\documentclass[journal]{IEEEtran}
\ifCLASSINFOpdf
\else
\fi
\usepackage{amsfonts,amssymb}
\usepackage{graphicx}
\usepackage{cite}
\usepackage{amssymb,amsfonts}
\usepackage{setspace}
\usepackage{textcomp}
\usepackage{color,soul}
\usepackage{multirow}
\usepackage{cuted}
\usepackage{array}
\usepackage{epsfig}
\usepackage{graphics}
\usepackage{varwidth}
\usepackage{amsmath}
\usepackage{breqn}
\usepackage{caption}
\usepackage{subcaption}
\usepackage{breqn}
\usepackage[T1]{fontenc}
\usepackage{mathtools}  
\usepackage{amsthm}
\usepackage{mathrsfs}
\usepackage{algorithm2e}
\usepackage{cleveref}
\allowdisplaybreaks
\usepackage{lipsum}
\newtheorem{remark}{Remark}
\begin{document}
%
\title{RIS-Aided Cooperative Mobile Edge Computing: Computation Efficiency Maximization via Joint Uplink and Downlink Resource Allocation}


\author{Zhenrong Liu,~\IEEEmembership{Graduate Student Member,~IEEE,}
        Zongze Li,\\
         Yi Gong,~\IEEEmembership{Senior Member,~IEEE,}
        and Yik-Chung~Wu,~\IEEEmembership{Senior Member,~IEEE}

}

%

\markboth{IEEE TRANSACTIONS ON WIRELESS COMMUNICATIONS,~Vol.~X, No.~X, X~2024}%
{Liu \MakeLowercase{\textit{et al.}}: RIS-Aided Cooperative Mobile Edge Computing: Computation Efficiency Maximization via Joint Uplink and Downlink Resource Allocation}

\IEEEpubid{0000--0000/00\$00.00~\copyright~2024 IEEE}


\onecolumn

{\fontsize{14}{16}\selectfont IEEE Copyright Notice}

\

\copyright~2024 IEEE. Personal use of this material is permitted. Permission from IEEE must be obtained for all
other uses, in any current or future media, including reprinting/republishing this material for advertising
or promotional purposes, creating new collective works for resale or redistribution to servers or lists, or
reuse of any copyrighted component of this work in other works.

\newpage
\twocolumn
\maketitle

\begin{abstract}
In mobile edge computing (MEC) systems, the wireless channel condition is a critical factor affecting both the communication power consumption and computation rate of the offloading tasks. This paper exploits the idea of cooperative transmission and employing reconfigurable intelligent surface (RIS) in MEC to improve the channel condition and maximize computation efficiency (CE). The resulting problem couples various wireless resources in both uplink and downlink, which calls for the joint design of the user association, receive/downlink beamforming vectors, transmit power of users, task partition strategies for local computing and offloading, and uplink/downlink phase shifts at the RIS. To tackle the challenges brought by the combinatorial optimization problem, the group sparsity structure of the beamforming vectors determined by user association is exploited. Furthermore, while the CE does not explicitly depend on the downlink phase shifts, instead of simply finding a feasible solution, we exploit the hidden relationship between them and convert this relationship into an explicit form for optimization. Then the resulting problem is solved via the alternating maximization framework, and the nonconvexity of each subproblem is handled individually. Simulation results show that cooperative transmission and RIS deployment can significantly improve the CE and demonstrate the importance of optimizing the downlink phase shifts with an explicit form.

\end{abstract}

\begin{IEEEkeywords}
Mobile edge computing (MEC), reconfigurable intelligent surface (RIS), computation efficiency, user association, cooperative transmission.
\end{IEEEkeywords}

%

\section{Introduction}
Mobile users have long been recognized as resource-limited compared to static clients and servers. At any given cost and technology level, weight, size, battery life, ergonomics, and heat dissipation severely limit computational resources in terms of processor speed, memory size, and storage capacity. On the other hand, a wide range of emerging computation-intensive applications, such as mobile virtual reality and augmented reality, call for an unprecedented demand for data computing and processing in mobile users. To liberate the resource-limited users from heavy computation workloads and provide them with high-performance computing services, mobile edge computing (MEC) promotes the use of computing capabilities at the edge servers attached to the wireless access points (APs) \cite{8016573,9439524}. In this way, users' computation-intensive tasks can be offloaded to the nearby APs, and the results are delivered back to the users after the computation is done at the edge servers. Unlike cloud computing, whose vision is to centralize computation, storage, and network management in a remote cloud center distant from the end users, MEC pushes computing resources, network control, and storage to the network edges\cite{9964436}. This dramatically reduces latency, mobile energy consumption, and communication overhead of the network backhaul, thus overcoming the key challenges for materializing the 5G vision.

As the focus in MEC shifts from pure communications to communication-assisted computation, it is important to consider optimizing the computation efficiency \cite{9380744}, which is the computing capability (in terms of computation rate) of the MEC system divided by power spent on computing and communications \cite{8986845}. However, increasing computing capability and decreasing power consumption are two conflicting goals. For example, if users are not offloading their tasks to the edge but computing them locally, power can be saved from communications, but the computation rate would be very low. On the contrary, if all users offload the tasks to the edge servers for computation, it causes high energy consumption for data offloading and results downloading, but the computation capability will be enhanced. Neither of these extremes is a good solution since they result in a waste of resources. Balancing the computation rate and the power consumption is a central problem in wireless MEC systems \cite{9018180,9270605,10120724}.

\IEEEpubidadjcol

A well-known strategy for reducing downlink transmission energy is to have multiple APs cooperatively serve each user \cite{4675744,8575160,8110665}. Different from the cellular systems, in wireless MEC, the data in downlink cooperative transmission is the computation results from users' offloaded data \cite{10091816}. To enable cooperative transmission, the APs must either share the computation results of their respective users or each AP compute the results independently. Unfortunately, the computation results exchange between APs alone would cost 15 ms (for a lightly loaded network) to 50 ms (for a heavily loaded network) \cite{8376975,6612630}, thus impractical for the latency-critical MEC applications such as mobile virtual reality and augmented reality, whose recommended total latency is around 20 ms \cite{raaen2015measuring}. Under the scenario that each cooperating AP independently computes the result for a serving user, more participating APs means more computation energy is spent in the hope of trading for less downlink transmission energy. This raises the fundamental question of which APs should be associated with which user. If the channel from a particular user $k$ to an AP $n$ is in bad condition, AP $n$ should not cooperatively serve the user $k$ since the offloading energy from user $k$ to AP $n$ will be huge, thus lowering the computation efficiency. But if we are too conservative in choosing APs for cooperation, it defeats the original purpose of improving the quality of downlink transmission. Therefore, user association is critical to the overall MEC system’s computation efficiency.


Besides cooperative transmission, another recent technology that could help to increase computation efficiency is reconfigurable intelligent surfaces (RISs). The motivation for incorporating RIS into MEC is driven by its remarkable capability to configure the wireless environment and thus improve the offloading rate and reduce transmission energy. Furthermore, RIS offers the advantage of being cost-effective and easy to deploy, making RIS particularly suitable for enhancing the uplink performance of MEC systems. By capitalizing on these benefits, the integration of RIS into the MEC framework holds great promise.


In this paper, we propose to use both RIS and cooperative transmission to improve the computation efficiency of an MEC system. In particular, we formulate the computation efficiency maximization problem by considering both the uplink and downlink transmission power consumption, and the local and offloaded computation power consumption. Since user association in the proposed MEC system couples with the design of receive and downlink beamforming vectors (i.e., if a user is not selected to be served by a particular AP, the corresponding receive and downlink beamforming vectors should be all-zero vectors), the resulting optimization problem is an intertwined design of user association, power partition parameters for computation offloading, receive and downlink beamforming vectors, and receive/downlink phase-shift matrices at the RIS with the practical phase shift model. Further with the fact that user association is a non-convex and combinatorial problem, the overall optimization problem is difficult to solve.

To address the challenge of the combinatorial optimization problem, we exploit the group sparsity structure of the receive/downlink beamforming vectors and merge the user association design into the beamforming variables. In addition, we observe that although the downlink phase-shift matrix does not appear in the objective function, it indirectly affects power consumption. This inspires us to exploit their hidden relationship to optimize the downlink phase-shift matrix with an explicit form instead of simply finding a feasible one. With the above two ideas, the optimization problem is solved under the alternating maximization (AM) framework. However, each of the subproblems is still non-convex. Tailoring to the different natures of non-convexity, these subproblems are handled by the tightest convex relaxation, Dinkelbach’s or quadratic transformation, and the penalty method, respectively. Extensive simulation results show that the proposed RIS-aided cooperative MEC system outperforms systems without RIS or cooperative transmission. At the same time, optimizing the downlink phase shifts with an explicit form can further improve the system's performance.

The rest of the paper is organized as follows. The system model and the computation efficiency maximization problem are formulated in Section \ref{sym}. The discrete user association variable is handled in Section \ref{3}. Section \ref{AM} maximizes the computation efficiency under the AM framework. Simulation results are presented in Section V. Finally, a conclusion is drawn in Section VI.

\textit{Notations}: We use boldface lowercase (e.g., $\boldsymbol{h}$) and uppercase letters (e.g., $\boldsymbol{G}$) to represent vectors and matrices, respectively. The transpose, conjugate, conjugate transpose, and diagonal matrix are denoted as $(\cdot)^{\mathrm{T}},(\cdot)^{*},(\cdot)^{\mathrm{H}}$ and $\operatorname{diag}(\cdot)$, respectively. The symbol $\Re(\cdot)$ denotes the real component of a complex number. The $n \times n$ identity matrix is denoted as $\boldsymbol{I}_{n}$ and the entries of a matrix $\boldsymbol{X}$ is denoted by $(\boldsymbol{X})_{i j}$. The complex normal distribution is denoted as $\mathcal{C N}$. The $\ell_{2}$-norm of a vector is denoted as $\|\cdot\|_{2}$ and the Frobenius norm of a matrix is denoted as $\|\cdot\|_{\text{F}}$. $[a]^{+} \text {denotes } \max (0, a)$. We use $\mathbf{1}_{\{\cdot\}}$ to denote the indicator function which outputs 1 if the condition $\{\cdot\}$ is satisfied and outputs 0 otherwise. In the rest of this paper, the superscripts U and D refer to uplink and downlink, respectively, and the letters d and r in the subscripts stand for the direct link and the reflected link, respectively. Finally, the set of a variable is denoted by \(\{variable\}\).

\section{System Model and Problem Formulation}
\label{sym}
\subsection{System Model}
\begin{figure}[t]
\centering
\centerline{\includegraphics[scale=0.08]{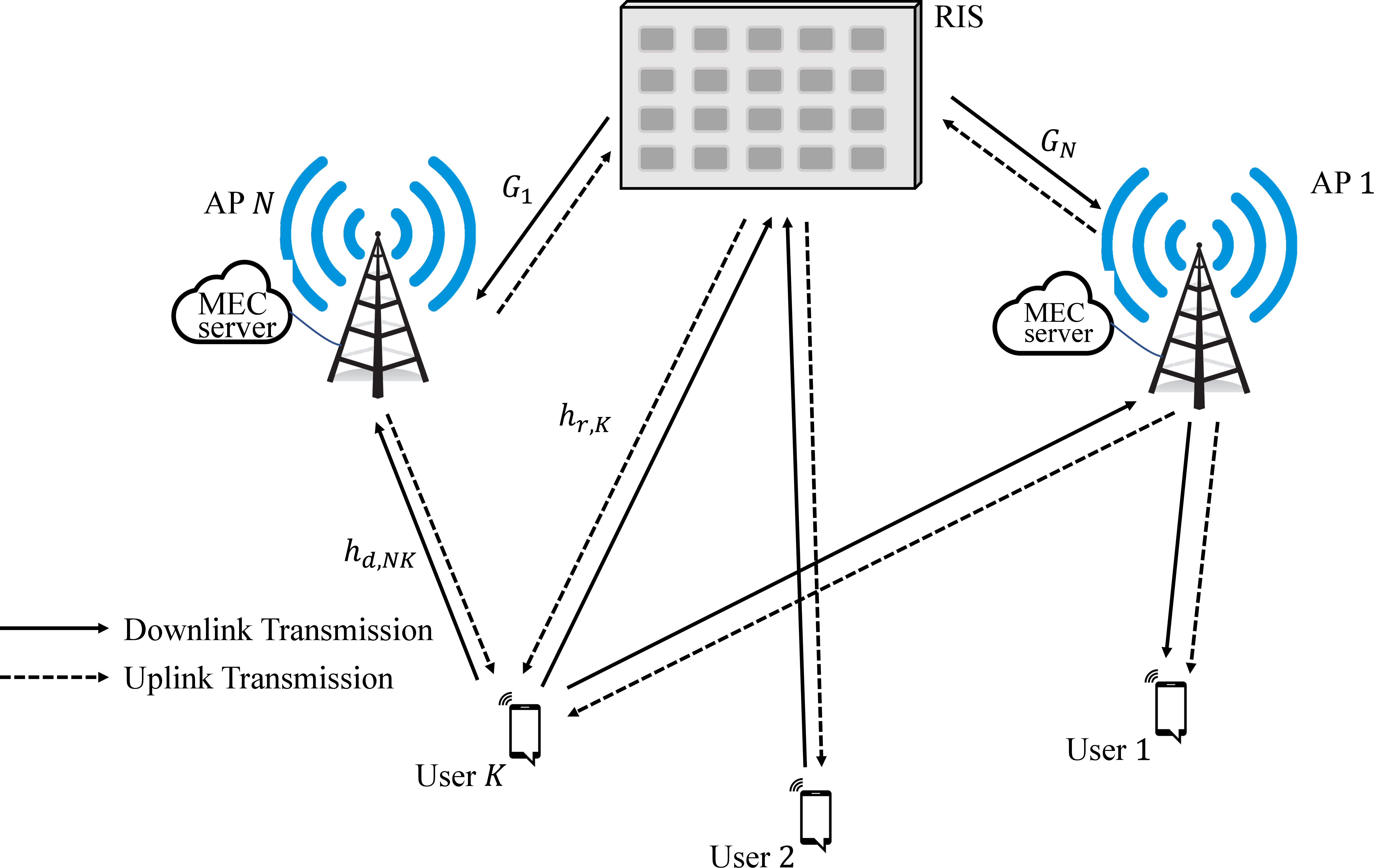}}
\caption{The RIS-aided MEC system.}
\label{sm}
\end{figure}

We consider a RIS-aided MEC system as shown in Fig. \ref{sm}, in which there are $N$ APs, $K$ single-antenna users, and an $M$-element RIS. Each AP is attached to an MEC edge server and equipped with $L$ antennas.
Each user has a limited power budget $P_{k}^{c}$ in Watt (W) but has intensive computation tasks to deal with. In practice, if the computation task is too complicated to be completed by a user (possibly due to excessive power or time involved), part of the task should be delegated to the edge server. Hence, a partial offloading mode is adopted. Partial offloading is designed to cope with computation tasks with data partition, in which tasks can be arbitrarily divided to facilitate parallel operations at users for local computing and offloading to the APs for edge computing\cite{7542156}. 
We use time division duplex protocol and adjust RIS phase shifts between uplink and downlink to reconfigure the propagation environment in real-time via a control link between RIS and APs\cite{9950554}. To guarantee the quality of services provided to users, it is assumed that multiple APs in the uplink transmission can successfully receive each user's data, enabling computation repetition at different edge servers and thus creating multiple copies of the computation results at different APs \cite{9062599}. These copies in turn enable cooperative downlink transmission among the APs on delivering the computation results.
By exploiting existing channel estimation approaches \cite{9087848}, the global channel state information is assumed to be available at the APs.

For uplink transmission, let $\mathcal{N}=\{1, \ldots, N\}$, $\mathcal{K}=\{1, \ldots, K\}$, and $\mathcal{M}=$ $\{1, \ldots, M\}$ denote the index sets of APs, users, and reflecting elements, respectively. In a specific time slot, let $s_{k}^{\mathrm{U}} \in \mathbb{C}$ with zero mean and unit variance denote the information symbol of user $k$ for the offloading task, and $p_{k}^{\mathrm{U}} \in \mathbb{R}$ denote the transmit power of user $k$. The signal received at AP $n$ can be expressed as
\begin{equation}
\boldsymbol{y}_{n}^{\mathrm{U}}=\sum_{k \in \mathcal{K}} \boldsymbol{\iota}_{n k}^{\mathrm{U}} \sqrt{p_{k}^{\mathrm{U}}} s_{k}^{\mathrm{U}}+\boldsymbol{z}_{n}^{\mathrm{U}}, \quad \forall n \in \mathcal{N},
\end{equation}
where $\boldsymbol{\iota}_{n k}^{\mathrm{U}} \in \mathbb{C}^{L \times 1}$ is the equivalent baseband channel from user $k$ to AP $n$ and $\boldsymbol{z}_{n}^{\mathrm{U}} \sim \mathcal{C N}\left(\mathbf{0}, \sigma_{n}^{2} \boldsymbol{I}_{L}\right)$ is
the receiver noise at AP $n$ with $\sigma_{n}^{2}$ being the noise power. With the deployment of an RIS, the equivalent baseband channel from user $k$ to AP $n$ consists of direct and reflected links. Therefore, $\boldsymbol{\iota}_{n k}^{\mathrm{U}}$ can be modeled as
\begin{equation}
\label{uplinkchannel}
\boldsymbol{\iota}_{n k}^{\mathrm{U}}=\boldsymbol{h}_{\mathrm{d}, n k}^{\mathrm{U}}+\left(\boldsymbol{G}_{n}^{\mathrm{U}}\right)^{\mathrm{H}}\boldsymbol{\Theta}^{\mathrm{U}} \boldsymbol{h}_{\mathrm{r}, k}^{\mathrm{U}}, \quad \forall k \in \mathcal{K}, n \in \mathcal{N},
\end{equation}
where $\boldsymbol{h}_{\mathrm{d}, n k}^{\mathrm{U}} \in \mathbb{C}^{L \times 1}, \boldsymbol{h}_{\mathrm{r}, k}^{\mathrm{U}} \in \mathbb{C}^{M \times 1}$, and $\boldsymbol{G}_{n}^{\mathrm{U}} \in \mathbb{C}^{M \times L}$ denote the channels from user $k$ to AP $n$, from user $k$ to the RIS, and from the RIS to AP $n$, respectively. The uplink RIS matrix $\boldsymbol{\Theta}^{\mathrm{U}}=\operatorname{diag}\left(\rho(\theta_{1}^{\mathrm{U}})e^{j\theta_{1}^{\mathrm{U}}}, \ldots, \rho(\theta_{M}^{\mathrm{U}})e^{j\theta_{M}^{\mathrm{U}}}\right) \in \mathbb{C}^{M \times M}$ is a diagonal matrix, where $\rho(\theta_{m}^{\mathrm{U}})$ is the amplitude reflection coefficient that is a function of uplink phase shift at the $m^{th}$ RIS element $\theta_m^{\mathrm{U}}\in [0, 2 \pi)$, and 
can be expressed as \cite{9115725}
\begin{equation}
\label{pf}
\rho\left(\theta_{m}^{\mathrm{U}}\right)=\left(1-\beta_{\min }\right)\!\!\left(\frac{\sin \left(\theta_{m}^{\mathrm{U}}-\phi\right)+1}{2}\right)^{\alpha}\!\!+\beta_{\min }, \  \forall m \in \mathcal{M},
\end{equation}
with $\beta_{\min } \geq 0, \phi \geq 0$, and $\alpha \geq 0$ are parameters related to the specific circuit implementation.

We consider the linear beamforming strategy and denote $\boldsymbol{v}_{n k}^{\mathrm{U}} \in$ $\mathbb{C}^{L \times 1}$ as the receive beamforming vector of AP $n$ for decoding $s_{k}^{\mathrm{U}}$. AP $n$ only decodes user $k$'s transmitted symbol $s_{k}^{\mathrm{U}}$ if it belongs to the set of AP with indices $\mathcal{A}_{k} \subseteq \mathcal{N}$, which are responsible for computing the partial computation tasks for user $k$. If $n \in \mathcal{A}_{k}$, the received signal at AP $n$ for user $k$, denoted by ${\hat{s}}_{n k}^{\mathrm{U}} \in \mathbb{C}$, is given by
\begin{equation}
\begin{aligned}
{\hat{s}}_{n k}^{\mathrm{U}}=&\left(\boldsymbol{v}_{n k}^{\mathrm{U}}\right)^{\mathrm{H}} {\boldsymbol{\iota}}_{n k}^{\mathrm{U}} \sqrt{p_{k}^{\mathrm{U}}} s_{k}^{\mathrm{U}}+\left(\boldsymbol{v}_{n k}^{\mathrm{U}}\right)^{\mathrm{H}}\!\!\sum_{{l \neq k, l \in \mathcal{K}}} {\boldsymbol{\iota}}_{n l}^{\mathrm{U}} \sqrt{p_{l}^{\mathrm{U}}}s_l^{\mathrm{U}}\\
&+\left(\boldsymbol{v}_{n k}^{\mathrm{U}}\right)^{\mathrm{H}} \boldsymbol{z}_{n}^{\mathrm{U}}, \quad \forall k \in \mathcal{K}.
\end{aligned}
\end{equation}

For the offloading task, we introduce a power partition parameter $a_{k} \in[0,1]$ for user $k$, and $p_{k}^{\mathrm{U}}=a_{k} P_{k}^{c}$ represents the power used for data offloading. Correspondingly, the uplink signal-to-interference-plus-noise ratio (SINR) observed at AP $n$ for user $k$ is thus given by
\begin{equation}
\label{upsinr}
\begin{aligned}
&\gamma_{n k}^{\mathrm{U}}\left(\{a_k\}, \boldsymbol{v}_{nk}^{\mathrm{U}},\{\theta_m^{\mathrm{U}}\}\right)\\
&=\!\!\frac{p_{k}^{\mathrm{U}}\left|\left(\boldsymbol{v}_{nk}^{\mathrm{U}}\right)^{\mathrm{H}}\!
\left(\boldsymbol{h}_{\mathrm{d}, n k}^{\mathrm{U}}\!+\!\left(\boldsymbol{G}_{n}^{\mathrm{U}}\right)^{\mathrm{H}}\!\!\boldsymbol{\Theta}^{\mathrm{U}} \boldsymbol{h}_{\mathrm{r}, k}^{\mathrm{U}}\right)
\right|^{2}}{\sum\limits_{l \neq k,l \in \mathcal{K}} p_{l}^{\mathrm{U}}\!\left|\left(\boldsymbol{v}_{n k}^{\mathrm{U}}\right)^{\mathrm{H}}\!\!
\left(\!\boldsymbol{h}_{\mathrm{d}, n l}^{\mathrm{U}}\!+\!\left(\boldsymbol{G}_{n}^{\mathrm{U}}\right)^{\mathrm{H}}\!\!\boldsymbol{\Theta}^{\mathrm{U}} \boldsymbol{h}_{\mathrm{r}, l}^{\mathrm{U}}\!\right)
\!\right|^{2}\!\!\!+\!\sigma_{n}^{2}\left\|\boldsymbol{v}_{n k}^{\mathrm{U}}\right\|_{2}^{2}},\\
&\quad \forall n \in \mathcal{A}_{k}, k \in \mathcal{K}.
\end{aligned}
\end{equation}

After performing the computation tasks, the APs cooperatively transmit the computation results to the corresponding users through downlink wireless channels. We assume the downlink transmission starts simultaneously from all serving APs in our cooperative transmission protocol\footnote{During the downlink phase, APs can employ alignment procedures to transmit only the finished bits across all the APs for a specific user at this time slot, and any remaining bits can be transmitted in the subsequent time slot’s downlink phase.}. Let $\{\mathcal{A}_{k}\}$ denote the task selection strategy, $s_{k}^{\mathrm{D}} \in \mathbb{C}$ with zero mean and unit variance denote the symbol intended for user $k$, and $\boldsymbol{v}_{n k}^{\mathrm{D}} \in \mathbb{C}^{L \times 1}$ denote the downlink beamforming vector from AP $n$ to user $k$.
The signal transmitted by AP $n$, denoted as $\boldsymbol{x}_{n}^{\mathrm{D}} \in \mathbb{C}^{L \times 1}$, is a summation of beamformed symbols for all users with $n \in \mathcal{A}_{k}$, i.e., $\boldsymbol{x}_{n}^{\mathrm{D}}=\sum_{\{k \mid  n \in \mathcal{A}_k, k \in \mathcal{K} \}} \boldsymbol{v}_{n k}^{\mathrm{D}} s_{k}^{\mathrm{D}}, \forall n \in \mathcal{N}$. Then, the signal received by user $k$ is expressed as
\begin{equation}
\begin{aligned}
y_{k}^{\mathrm{D}}&=\sum_{n \in \mathcal{N}} \mathbf{1}_{\left\{n \in \mathcal{A}_{k}\right\}}\left({\boldsymbol{\iota}}_{n k}^{\mathrm{D}}\right)^{\mathrm{H}} \boldsymbol{v}_{n k}^{\mathrm{D}} s_{k}^{\mathrm{D}}\\
&+\sum_{l \neq k} \sum_{n \in \mathcal{N}} \mathbf{1}_{\left\{n \in \mathcal{A}_{l}\right\}}\left({\boldsymbol{\iota}}_{n k}^{\mathrm{D}}\right)^{\mathrm{H}} \boldsymbol{v}_{n l}^{\mathrm{D}} s_{l}^{\mathrm{D}}+z_{k}^{\mathrm{D}}, \quad \forall k \in \mathcal{K},
\end{aligned}
\end{equation}
where $z_{k}^{\mathrm{D}} \in \mathbb{C}$ is the AWGN at user $k$ with zero mean and variance $\sigma_{k}^{2}$, and $\boldsymbol{\iota}_{n k}^{\mathrm{D}} \in \mathbb{C}^{L\times 1}$ is the equivalent baseband channel from AP $n$ to user $k$. Similar to (\ref{uplinkchannel}), $\boldsymbol{\iota}_{nk}^{\mathrm{D}}$ is expressed as
\begin{equation}
\label{downlinkchannel}
\boldsymbol{\iota}_{n k}^{\mathrm{D}}=\boldsymbol{h}_{\mathrm{d}, n k}^{\mathrm{D}}+\left(\boldsymbol{G}_{n}^{\mathrm{D}}\right)^{\mathrm{H}}\boldsymbol{\Theta}^{\mathrm{D}} \boldsymbol{h}_{\mathrm{r}, k}^{\mathrm{D}}, \quad \forall k \in \mathcal{K}, n \in \mathcal{N},
\end{equation}
where $\boldsymbol{h}_{\mathrm{d}, n k}^{\mathrm{D}} \in \mathbb{C}^{L\times 1},\boldsymbol{h}_{\mathrm{r}, k}^{\mathrm{D}} \in \mathbb{C}^{M \times 1}$, and $\boldsymbol{G}_{n}^{\mathrm{D}} \in$
$\mathbb{C}^{M \times L}$ denote the channels from AP $n$ to user $k$, from the RIS to user $k$, and from $\mathrm{AP}$ $n$ to the RIS, respectively. Furthermore, $\boldsymbol{\Theta}^{\mathrm{D}}=\operatorname{diag}\left(\rho(\theta_{1}^{\mathrm{D}})e^{j\theta_{1}^{\mathrm{D}}}, \ldots, \rho(\theta_{M}^{\mathrm{D}})e^{j\theta_{M}^{\mathrm{D}}}\right) \in \mathbb{C}^{M \times M}$ is the downlink RIS matrix with $\theta_{m}^{\mathrm{D}} \in[0,2 \pi)$. Based on the above equation, the SINR observed by user $k \in \mathcal{K}$ in the downlink transmission is given by
\begin{align}
&\text{SINR}_k^{\mathrm{D}}\left(\{{\boldsymbol{v}_{nk}^{\mathrm{D}}}\}, \{\theta_m^{\mathrm{D}}\}\right) \nonumber\\
&=\!\frac{\left|\sum_{n \in \mathcal{N}} \mathbf{1}_{\left\{n \in \mathcal{A}_{k}\right\}}
\!\!\left(\!\boldsymbol{h}_{\mathrm{d}, n k}^{\mathrm{D}}\!+\!\left(\boldsymbol{G}_{n}^{\mathrm{D}}\right)^{\mathrm{H}}\!\boldsymbol{\Theta}^{\mathrm{D}}\boldsymbol{h}_{\mathrm{r}, k}^{\mathrm{D}}\!\right)^{\mathrm{H}}\!\! \boldsymbol{v}_{n k}^{\mathrm{D}}\right|^{2}}{\sum_{l \neq k}\!\left|\sum_{n \in \mathcal{N}} \!\mathbf{1}_{\left\{n \in \mathcal{A}_{l}\right\}}\!\!\left(\!\boldsymbol{h}_{\mathrm{d}, n k}^{\mathrm{D}}\!+\!\left(\boldsymbol{G}_{n}^{\mathrm{D}}\right)^{\mathrm{H}}\!\!\boldsymbol{\Theta}^{\mathrm{D}} \boldsymbol{h}_{\mathrm{r}, k}^{\mathrm{D}}\!\right)^{\mathrm{H}}\!\! \!\!\boldsymbol{v}_{n l}^{\mathrm{D}}\right|^{2}\!\!\!+\!\sigma_{k}^{2}}, \nonumber\\
&=\frac{\left|\sum_{n \in \mathcal{N}}\!\left(\!\boldsymbol{h}_{\mathrm{d}, n k}^{\mathrm{D}}\!+\!\left(\boldsymbol{G}_{n}^{\mathrm{D}}\right)^{\mathrm{H}}\!\boldsymbol{\Theta}^{\mathrm{D}}\boldsymbol{h}_{\mathrm{r}, k}^{\mathrm{D}}\!\right)^{\mathrm{H}}\!\!\boldsymbol{v}_{nk}^{\mathrm{D}}\right|^{2}}{\sum_{l \neq k}\left|\sum_{n \in \mathcal{N}}\!\left(\!\boldsymbol{h}_{\mathrm{d}, n k}^{\mathrm{D}}\!+\!\left(\boldsymbol{G}_{n}^{\mathrm{D}}\right)^{\mathrm{H}}\!\boldsymbol{\Theta}^{\mathrm{D}} \boldsymbol{h}_{\mathrm{r}, k}^{\mathrm{D}}\!\right)^{\mathrm{H}}\!\! \boldsymbol{v}_{n l}^{\mathrm{D}}\right|^{2}+\sigma_{k}^{2}},
\end{align}
where the second equality holds because AP $n$ does not transmit data to user $k$ by setting $\boldsymbol{v}_{n k}^{\mathrm{D}}=\mathbf{0}$ if $n \notin \mathcal{A}_{k}$.
\begin{remark}
Given the coverage area of RIS in an experimental system \cite{9551980}, as well as the density/coverage area of the base station in 5G NR \cite{ericsson2023,qualcomm2023}, utilizing a single RIS for multiple MEC servers is possible in practical deployments. This assumption is also widely accepted in RIS-aided MEC research\cite{9352968,10221786}.
\end{remark}
\begin{remark}
The RIS model described by (\ref{pf}) can be switched to the simultaneously transmitting and reflecting (STAR-) RIS model, thus enabling serving both users at the transmission and reflection side\cite{9570143}. The resulting problem is simpler than the adopted practical phase shift model since the amplitude coefficients and phase shifts of the STAR-RIS are not coupled.
\end{remark}
\subsection{Offloading Model}
In our system model, the communication rate of user $k$ offloaded to AP $n$ is
\begin{equation}
\label{upsm}
\begin{aligned}
R_{nk}\left(\{a_k\}, \boldsymbol{v}_{nk}^{\mathrm{U}},\{\theta_m^{\mathrm{U}}\}\right)\!=&B\log _{2}\left(1+\gamma_{n k}^{\mathrm{U}}\left(\{a_k\}, \boldsymbol{v}_{nk}^{\mathrm{U}},\{\theta_m^{\mathrm{U}}\}\right)\right),\\
&\quad \forall k \in \mathcal{K}, n \in \mathcal{N},
\end{aligned}
\end{equation}
where $B$ is the bandwidth of the system. 
However, due to users offloading to multiple APs through multicast transmission, for user $k$, its communication rate is determined by the slowest rate among all the APs they are connected to. This is mathematically expressed as $\min_{n \in \mathcal{A}_k} R_{nk}.$ Each user is allocated a data transmission time $t_k$, and the average computation rate for user $k$ is expressed as $(t_k\min_{n \in \mathcal{A}_k} R_{nk})/ T$, where $T$ represents the time slot length. After data uploading to AP $n$, a computation frequency $f_{nk}$ is allocated to each user $k$. If $C_k$ is the amount of required computing resource (i.e., the number of CPU cycles) for computing 1-bit of the user $k$'s input data, then we can calculate the maximum computation time among all the connected APs for user $k$'s task as $(C_kt_k\min_{n \in \mathcal{A}_k} R_{nk})/\min_{n \in \mathcal{A}_k}f_{nk}$.

As for the case of local computing, the dynamic voltage and frequency scaling technique is adopted by all users for increasing the computation energy efficiency through adaptively controlling the CPU frequency used for computing \cite{6574874}.
In particular, the computation energy consumption of user $k \in \mathcal{K}$ can be expressed as $T\kappa_kf_k^2$, where $f_k$ is the local computation frequency, $\kappa_k$ is the effective capacitance coefficient of user $k$. Also, as $(1-a_k)P_k^c$ is the allocated power for local computing, we have $T(1-a_k)P_k^c=T\kappa_kf_k^2$, and thus we can calculate $f_k$ as $f_{k}=\sqrt{\left((1-a_k)P_{k}^{c}\right)/\kappa_k},\ \forall k \in \mathcal{K}$. The computation rate of user $k$ for local computing is then given by
\begin{equation}
\label{lcr}
R_{k}^{\mathrm{loc}}\left(a_{k}\right)=\frac{f_{k} }{C_{k}}=\frac{1}{C_{k}} \sqrt{\frac{(1-a_k)P_{k}^{c}}{\kappa_k}}, \quad \forall k \in \mathcal{K}.
\end{equation}

\subsection{Power Consumption and Latency Model}

The power consumption of each user's computation task includes both that used in local computing and computation at APs for offloaded tasks. To be specific, at user $k$, $\left(1-a_{k}\right)P_{k}^{c}$ W will be used for local computing\cite{8334188}. In terms of the edge computation at the AP $n$, the average computation power is given as $\left((C_kt_k\min_{n \in \mathcal{A}_k} R_{nk}/f_{nk})f_{nk}^2\kappa_n\right)/T$. Therefore, the total computation power consumption for all APs and users is given by
$\sum_{k \in \mathcal{K}}\sum_{n \in \mathcal{A}_k} \left(C_kt_kf_{nk}\kappa_n\min_{n \in \mathcal{A}_k} R_{nk}\right)/T+\sum_{k \in \mathcal{K}}(1-a_k)P_{k}^{c}$.

In terms of communication power consumption, it consists of the power consumed by the users in the data offloading and APs in the downlink transmission. The total data offloading power consumption is $\sum_{k \in \mathcal{K}} p_{k}^{\mathrm{U}}=\sum_{k \in \mathcal{K}}a_{k} P_{k}^{c}$, while the downlink transmit power consumption from all the APs to a specific user $k$ is represented by $\sum_{n \in \mathcal{A}_{k}}\left\|\boldsymbol{v}_{n k}^{\mathrm{D}}\right\|_{2}^{2}$.
Therefore, the total communication power consumption for both uplink and downlink transmissions is given by $\sum_{k \in \mathcal{K}}a_{k} P_{k}^{c}+\sum_{k \in \mathcal{K}} \sum_{n \in \mathcal{A}_{k}}\left\|\boldsymbol{v}_{n k}^{\mathrm{D}}\right\|_{2}^{2}$.
Combining both computation and communication power consumption, the overall system power consumption can be expressed as
\begin{equation}
\label{power}
\begin{aligned}
&P_{\text{total}}\!\left(\{\mathcal{A}_k\},\!\left\{\boldsymbol{v}_{n k}^{\mathrm{D}}\right\}\!,\{\boldsymbol{v}_{nk}^{\mathrm{U}}\},\{f_{nk}\},\{a_k\},\{t_k\},\{\theta_m^{\mathrm{U}}\}\right)\!= \!\!\sum_{k \in \mathcal{K}}\!P_{k}^{c}\\
&+\sum_{k \in \mathcal{K}} \sum_{n \in \mathcal{A}_{k}}\left\|\boldsymbol{v}_{n k}^{\mathrm{D}}\right\|_{2}^{2} + \!\!\sum_{k \in \mathcal{K}}\!\sum_{n \in \mathcal{A}_{k}}\!\!\!\frac{\left(C_kt_kf_{nk}\kappa_n\min_{n \in \mathcal{A}_k} R_{nk}\right)}{T}.
\end{aligned}
\end{equation}

Assuming the minimum required task’s data size for user $k$ is denoted as $U_k$ bits \cite{9812481}. Since local computing is ongoing during the entire time slot, the number of computed bits is $TR_k^{\text{loc}}$ \cite{8986845}, and the remaining data bits for offloading is $U_k -TR_k^{\text{loc}}$. Then the offloading latency for user $k$ plus computation latency at AP $n$ is calculated as 
\begin{equation}
\label{12}
\tau_{nk}=(U_k -TR_k^{\text{loc}})/\min_{n \in \mathcal{A}_k} R_{nk} + (U_k - TR_k^{\text{loc}})/f_{nk},
\end{equation}
where the first term represents data offloading latency, and the second term accounts for computation latency at the AP.

\subsection{Problem Formulation}
\begin{figure*}[ht]
\begin{subequations}
\label{originorigin}
\begin{align}
\mathtt{P}_{0}\!:\!\max _{\substack{\{\boldsymbol{v}_{nk}^{\mathrm{U}}\},\{\boldsymbol{v}_{nk}^{\mathrm{D}}\}, \{\mathcal{A}_{k}\}, \{f_{nk}\},\\ \{\theta_m^{\mathrm{U}}\},\{\theta_m^{\mathrm{D}}\}, \{a_k\},\{t_k\}}} 
\quad &\frac{\sum_{k=1}^{K}{t_k}\min_{n \in \mathcal{A}_k} R_{nk}\left(\{a_k\}, \boldsymbol{v}_{nk}^{\mathrm{U}},\{\theta_m^{\mathrm{U}}\}\right)/{T}+\sum_{k=1}^{K}R_{k}^{\mathrm{loc}}\left(a_{k}\right)}{P_{\text{total}}\!\left(\{\mathcal{A}_k\},\!\left\{\boldsymbol{v}_{n k}^{\mathrm{D}}\right\}\!,\{\boldsymbol{v}_{nk}^{\mathrm{U}}\},\{f_{nk}\},\{a_k\},\{t_k\},\{\theta_m^{\mathrm{U}}\}\right)},\\
\label{a}
\text { s.t. } \quad \quad & a_{k} \in[0,1], \quad \forall k \in \mathcal{K}, \\
\label{latency}
&\tau_{nk} \leq \chi, \quad \forall k \in \mathcal{K}, \ \forall n \in \mathcal{A}_k,\\
\label{b}
&\mathbf{1}_{\left\{a_k \neq 0\right\}} \text{SINR}_k^{\mathrm{D}}\left(\{{\boldsymbol{v}_{nk}^{\mathrm{D}}}\}, \{\theta_m^{\mathrm{D}}\}\right) \geq \mathbf{1}_{\left\{a_k \neq 0\right\}} \gamma_{k}^{\mathrm{D}}, \quad k \in \mathcal{K},\\
\label{d}
&\sum_{k \in \mathcal{K}}\left\|\boldsymbol{v}_{n k}^{\mathrm{D}}\right\|_{2}^{2} \leq P_{n, \max }^{\mathrm{D}},\ \  \forall n \in \mathcal{N},\\
\label{g1}
&\boldsymbol{v}_{n k}^{\mathrm{U}}=\boldsymbol{v}_{n k}^{\mathrm{D}}=\mathbf{0}, \quad \forall n \notin \mathcal{A}_{k}, k \in \mathcal{K},\\
\label{UU}
& \boldsymbol{\Theta}^{\mathrm{U}/\mathrm{D}}=\operatorname{diag}\left(\rho(\theta_{1}^{\mathrm{U}/\mathrm{D}})e^{j\theta_{1}^{\mathrm{U}/\mathrm{D}}}, \ldots, \rho(\theta_{M}^{\mathrm{U}/\mathrm{D}})e^{j\theta_{M}^{\mathrm{U}/\mathrm{D}}}\right), \  \forall m \in \mathcal{M},\\
\label{tk}
&t_k+(C_kt_k\min_{n \in \mathcal{A}_k} R_{nk})/\min_{n \in \mathcal{A}_k}f_{nk} \leq T, \quad \forall k \in \mathcal{K},\\
\label{frequency}
&\sum_{k \in \mathcal{K}}f_{nk}\leq f_n, \ f_{nk} \geq 0,\quad \forall k \in \mathcal{K},\forall n \in \mathcal{N}.
\end{align}
\end{subequations}
\hrule  
\end{figure*}
In the proposed RIS-aided MEC system, there exists a fundamental tradeoff between the quality of service and the system power consumption. Specifically, with computation replication, more APs performing the same task improves the downlink diversity gain by exploiting cooperative transmission, at the cost of increasing the computation power consumption. To strike a good balance between communication power consumption, computation power consumption, and computation rate, we aim to maximize computation efficiency, defined as the ratio of total computation rate to total power consumption. This leads to the optimization problem being formulated as (\ref{originorigin}). 

In the formulated problem, (\ref{latency}) indicates that for each user $k$, the latency among all APs serving that user must be less than a threshold $\chi < T$. 
(\ref{b}) implies that for user $k$ to perform uplink offloading, the subsequent downlink SINR must exceed the specified threshold $\gamma_k^{\mathrm{D}}$ for reliable reception of computation results. An indicator function exists in (\ref{b}) since if the user is determined not to offload, the downlink SINR does not matter. In this case, both sides of the constraint become zero, and the constraint is satisfied automatically. $P_{n, \max }^{\mathrm{D}}$ in (\ref{d}) denote the maximum transmit power of AP $n$ in the downlink. (\ref{g1}) specifies that if $n \notin \mathcal{A}_{k}$, AP $n$ does not decode user $k$'s data in the uplink (i.e., $\boldsymbol{v}_{n k}^{\mathrm{U}}=\mathbf{0}$) and subsequently would not transmit computation results to user $k$ in the downlink (i.e., $\boldsymbol{v}_{n k}^{\mathrm{D}}=\mathbf{0}$). (\ref{g1}) reveals that the full set of the beamforming vectors $\{\boldsymbol{v}_{1 1}^{\mathrm{U}},\cdots,\boldsymbol{v}_{N K}^{\mathrm{U}},\boldsymbol{v}_{1 1}^{\mathrm{D}},\cdots,\boldsymbol{v}_{N K}^{\mathrm{D}}\}$ has a special structure where some components in the set are simultaneously zero. Constraint (\ref{tk}) ensures the user's data transmission and computation time do not exceed the time slot length. Further, (\ref{frequency}) limits the cumulative computational frequency $f_{nk}$ for each user $k$ to not surpass AP $n$'s total computational frequency $f_n$ \cite{9812481}. Notice that in the formulated problem, the norm $\|\boldsymbol{v}_{nk}^{\mathrm{U}}\|$ does not matter as $\boldsymbol{v}_{nk}^{\mathrm{U}}$ only appears in the objective function through $R_{nk}$, and from (\ref{upsinr}), $\boldsymbol{v}_{nk}^{\mathrm{U}}$ appears in both the numerator and denominator of the SINR expression.

Efficiently solving $\mathtt{P}_0$ is a highly challenging endeavor. Firstly, $\mathtt{P}_{0}$ is intricate due to its general non-convexity in both the objective function and constraints, compounded by the non-differentiability from the maximin optimization structure within $\min_{n \in \mathcal{A}_k} R_{nk}$ and $\min_{n \in \mathcal{A}_k}f_{nk}$. Secondly, since discrete variables are involved in $\{\mathcal{A}_k\}$, this makes $\mathtt{P}_0$ a combinatorial optimization problem. While brute force and randomized search algorithms are widely accepted techniques for solving the combinatorial optimization problem, they are not desirable solutions as brute force search is computationally prohibitive due to the exponential complexity\cite{yang2020sparse} while randomized search lacks solution quality guarantee \cite{lovasz1995randomized,neumann2006combinatorial}. 
Thirdly, since variable $\{\theta_m^{\mathrm{D}}\}$ does not appear in the objective function but indirectly affects the objective value through (\ref{b}), it is nontrivial to derive an optimal solution for $\{\theta_m^{\mathrm{D}}\}$.
In the subsequent sections, we begin by transforming the maximin optimization problem into a more tractable form. Following that, we leverage on the distinctive structure of the beamforming vectors to address the discrete variable $\{\mathcal{A}_k\}$ within $\mathtt{P}_{0}$, thus simplifying the subsequent algorithm design.
\begin{remark}
    In the proposed system, we assume that users determine the parameter $\gamma_k^{\mathrm{D}}$ based on their specific requirements and is independent of the optimization variable $a_k$ as in \cite{9234651,9286485}. This is because the number of bits uploaded to the AP does not directly relate to the number of bits in the downlink in wireless MEC for applications such as image classification and video tracking \cite{9031741}.
\end{remark}
\section{Transformation of Maximin Optimization and Handling Discrete Variable in $\mathtt{P}_{0}$}
\label{3}
The objective function of $\mathtt{P}_{0}$ is non-differentiable due to the maximin optimization structure, which is embedded in the terms $\min_{n \in \mathcal{A}_k} R_{nk}$ in the objective function and the terms $\min_{n \in \mathcal{A}_k} R_{nk}$ and $\min_{n \in \mathcal{A}_k}f_{nk}$ within the constraints (\ref{tk}). To address this, we reformulate it as follows:
\begin{subequations}
\begin{align}
&\mathtt{P}_{1}:\nonumber\\
&\!\!\!\!\max_{\substack{\{\boldsymbol{v}_{nk}^{\mathrm{U}}\},\{\boldsymbol{v}_{nk}^{\mathrm{D}}\}, \{f_{nk}\},\\ \{\mathcal{A}_{k}\},\{\theta_m^{\mathrm{U}}\}, \{\theta_m^{\mathrm{D}}\},\{t_k\}\\ \{R_k\},\{a_k\}, \{f_{k}^{\min}\}}}
\frac{\sum_{k=1}^{K}t_kR_k/T\!+\!\sum_{k=1}^{K}R_{k}^{\mathrm{loc}}\left(a_{k}\right)}{P_{\text {total }}\!\!\!\left(\{\mathcal{A}_k\},\{R_k\}, \left\{\boldsymbol{v}_{n k}^{\mathrm{D}}\right\},\{f_{nk}\},\{t_k\}\right)},\\
&\quad \text { s.t. }(\text{\ref{a}})-(\text{\ref{UU}}),(\text{\ref{frequency}}),\nonumber\\
\label{tk1}
&\quad\quad\quad t_k+(C_kt_kR_k)/f_k^{\min} \leq T, \quad \forall k \in \mathcal{K},\\
\label{maximinmin}
&\quad\quad\quad  R_k \leq \min_{n \in \mathcal{A}_k} R_{nk}\left(\{a_k\}, \boldsymbol{v}_{nk}^{\mathrm{U}},\{\theta_m^{\mathrm{U}}\}\right), \  \forall k \in \mathcal{K},\\
\label{maximinfrequency}
&\quad\quad\quad  f_k^{\min}\leq\min_{n \in \mathcal{A}_k}f_{nk}, \quad \forall k \in \mathcal{K},
\end{align}
\end{subequations}
where \(R_k\) and \(f_k^{\min}\) are introduced as auxiliary variables. Note that $\mathtt{P}_{1}$ and $\mathtt{P}_{0}$ are considered equivalent since to maximize the objective function, the value of $R_k$ and $f_k^{\min}$ needs to be set equal to $\min_{n \in \mathcal{A}_k} R_{nk}\left(\{a_k\}, \boldsymbol{v}_{nk}^{\mathrm{U}},\{\theta_m^{\mathrm{U}}\}\right)$ and $\min_{n \in \mathcal{A}_k}f_{nk}$, respectively\cite{visotsky1999optimum,9435051}.

Regarding the discrete variable $\{\mathcal{A}_k\}$, notice that it has an intrinsic connection with the group sparsity structure of the beamforming vectors. Specifically, $\boldsymbol{v}_{nk}^{\mathrm{U}}$ and $\boldsymbol{v}_{nk}^{\mathrm{D}}$ will be zero at the same time if $n \notin \mathcal{A}_{k}$ or non-zero at the same time if $n \in \mathcal{A}_{k}$. This inspires us to impose a value restriction on the mixed $\ell_{1,2}$ norm [35],[38]: $\sum_{n=1}^{N} \sum_{k=1}^{K} \left|\|[(\boldsymbol{v}_{n k}^{\mathrm{D}})^{\mathrm{T}}\ (\boldsymbol{v}_{n k}^{\mathrm{U}})^{\mathrm{T}}]\|_2\right|$ to enforce the group sparsity structure. In this formulation, the outer $\ell_{1}$-norm induces sparsity, while the inner $\ell_{2}$-norm is responsible for forcing all coefficients in the beamforming group $[(\boldsymbol{v}_{n k}^{\mathrm{D}})^{\mathrm{T}}\ (\boldsymbol{v}_{n k}^{\mathrm{U}})^{\mathrm{T}}]$ to be zero. By doing so, we do not have to explicitly optimize the task selection strategy $\{\mathcal{A}_k\}$. Instead, $\{\mathcal{A}_k\}$ can be determined by the group sparsity pattern of the beamforming vectors, i.e., $\mathcal{A}_{k}=\left\{n \mid [(\boldsymbol{v}_{n k}^{\mathrm{D}})^{\mathrm{T}}\ (\boldsymbol{v}_{n k}^{\mathrm{U}})^{\mathrm{T}}] \neq \mathbf{0}, n \in \mathcal{N}\right\}$.

Next, to leverage the sparsity structure for more efficient problem-solving, we need to transform the constraints (\ref{latency}), (\ref{maximinmin}) and (\ref{maximinfrequency}), which involve the variable $\{\mathcal{A}_k\}$. The equivalent transformation for the latency constraints (\ref{latency}) is given by
\begin{equation}
\label{latencynorm}
\begin{aligned}
        \mathbf{1}_{\left\{\boldsymbol{v}_{nk}^{\mathrm{D}} \neq \mathbf{0}\right\}}\!\Biggl(\frac{U_k -TR_k^{\text{loc}}}{R_k}\!+\!\frac{U_k -TR_k^{\text{loc}}}{f_{nk}} \Biggl)\!&\leq \!\mathbf{1}_{\left\{\boldsymbol{v}_{nk}^{\mathrm{D}} \neq \mathbf{0}\right\}} \chi, \\
        \quad & k \in \mathcal{K},  n \in \mathcal{N},
\end{aligned}
\end{equation}
which is obtained by putting the expression for $\tau_{nk}$ from equation (\ref{12}) into (\ref{latency}). In (\ref{latencynorm}), the indicator function signifies that for user $k$, if AP $n$ is not in $\mathcal{A}_k$, its latency is disregarded (as both sides of the inequality in (\ref{latencynorm}) become 0, automatically satisfying the constraint). Similarly, for (\ref{maximinmin}) and (\ref{maximinfrequency}), we can equivalently transform them as
\begin{equation}
\label{normmaximin}
   \mathbf{1}_{\left\{\boldsymbol{v}_{nk}^{\mathrm{D}} \neq \mathbf{0}\right\}} R_k \leq \mathbf{1}_{\left\{\boldsymbol{v}_{nk}^{\mathrm{D}} \neq \mathbf{0}\right\}} R_{nk}, \quad k \in \mathcal{K}, n \in \mathcal{N},
\end{equation}
and
\begin{equation}
\label{freqnorm}
    \mathbf{1}_{\left\{\boldsymbol{v}_{nk}^{\mathrm{D}} \neq \mathbf{0}\right\}}f_k^{\min}\leq\mathbf{1}_{\left\{\boldsymbol{v}_{nk}^{\mathrm{D}} \neq \mathbf{0}\right\}}f_{nk}, \quad k \in \mathcal{K}, n \in \mathcal{N},
\end{equation}
respectively.
Therefore, the transformed problem of $\mathtt{P}_{1}$ is given by
\begin{subequations}
\label{origin_constrained}
\begin{align}
\!\!\!\!\mathtt{P}_{2}:
&\max_{\substack{\{\boldsymbol{v}_{nk}^{\mathrm{U}}\},\{\boldsymbol{v}_{nk}^{\mathrm{D}}\},\{R_k\},\\ \{f_{nk}\},\{\theta_m^{\mathrm{U}}\},  \{\theta_m^{\mathrm{D}}\},\\ \{a_k\},\{f_{k}^{\min}\},\{t_k\}}} 
\frac{\sum_{k=1}^{K}\!t_kR_k/T\!+\!\sum_{k=1}^{K}R_{k}^{\mathrm{loc}}\left(a_{k}\right)}{P_{\text {total }}\!\!\!\left(\{R_k\}, \left\{\boldsymbol{v}_{n k}^{\mathrm{D}}\right\},\{f_{nk}\},\{t_k\}\right)}, \\
&\text { s.t. } \quad\quad (\text{\ref{a}}),(\text{\ref{b}}),(\text{\ref{d}}),
(\text{\ref{UU}}),\nonumber\\
&\quad\quad\quad\quad(\text{\ref{frequency}}),(\text{\ref{tk1}}),(\text{\ref{latencynorm}}),(\text{\ref{normmaximin}}),(\text{\ref{freqnorm}}),\nonumber\\
\label{pn}
&\quad\quad\quad\quad\sum_{n=1}^{N} \sum_{k=1}^{K} \left|\|[(\boldsymbol{v}_{n k}^{\mathrm{D}})^{\mathrm{T}}\ (\boldsymbol{v}_{n k}^{\mathrm{U}})^{\mathrm{T}}]\|_2\right| \leq \beta
\end{align}
\end{subequations}
for some $\beta > 0$ and $P_{\text {total }}\!\!\!\left(\{R_k\}, \left\{\boldsymbol{v}_{n k}^{\mathrm{D}}\right\},\{f_{nk}\},\{t_k\}\right)$ can be equivalently rewritten as 
\begin{equation}
\label{denominator_without}
\begin{aligned}
&P_{\text {total }}\!\!\!\left(\{R_k\}, \left\{\boldsymbol{v}_{n k}^{\mathrm{D}}\right\},\{f_{nk}\},\{t_k\}\right)=\sum_{k \in \mathcal{K}}P_{k}^{c}\\
&+\sum_{n=1}^{N}\sum_{k=1}^{K} \mathbf{1}_{\left\{\boldsymbol{v}_{nk}^{\mathrm{D}} \neq \mathbf{0}\right\}} \left(\left\|\boldsymbol{v}_{n k}^{\mathrm{D}}\right\|_{2}^{2}+\frac{\left(C_kt_kR_{k}f_{nk}\kappa_n\right)}{T}\right),
\end{aligned}
\end{equation}
where an indicator function is used in the second term due to a change from summation over $\mathcal{A}_k$ to summation over all $n$.

Finally, to address the non-convex constraints (\ref{normmaximin}), we employ the logarithmic barrier function $\frac{1}{w} \log (-x)$ to replace the inequality and move it as a penalty term
\begin{align}
\begin{split}
&\mathtt{P}_{2w}:\max_{\substack{\{\boldsymbol{v}_{nk}^{\mathrm{U}}\},\{\boldsymbol{v}_{nk}^{\mathrm{D}}\}, \{R_k\},\\ \{f_{nk}\},\{\theta_m^{\mathrm{U}}\},  \{\theta_m^{\mathrm{D}}\},\\ \{a_k\},\{f_{k}^{\min}\},\{t_k\}}}\frac{\sum_{k=1}^{K}\!t_kR_k/T\!+\!\sum_{k=1}^{K}R_{k}^{\mathrm{loc}}\left(a_{k}\right)}{P_{\text {total }}\!\!\!\left(\{R_k\}, \left\{\boldsymbol{v}_{n k}^{\mathrm{D}}\right\},\{f_{nk}\},\{t_k\}\right)} \\
&\quad\quad\quad\quad\quad+\frac{1}{w}\sum_k\sum_n\mathbf{1}_{\left\{\boldsymbol{v}_{nk}^{\mathrm{D}} \neq \mathbf{0}\right\}}\left(\log( R_{nk}- R_k)\right),
\end{split}\\
&\text { s.t. } \quad \quad 
(\text{\ref{a}}),(\text{\ref{b}}),(\text{\ref{d}}),
(\text{\ref{UU}}),\nonumber\\
&\quad\quad\quad\ \ \ \!(\text{\ref{frequency}}),(\text{\ref{tk1}}),(\text{\ref{latencynorm}}),(\text{\ref{freqnorm}}),(\text{\ref{pn}}).\nonumber
\end{align}
As $w$ increases towards infinity, $\mathtt{P}_{2w}$ becomes equivalent to $\mathtt{P}_{2}$ \cite{9435051}. Therefore, we implement a strategy for solving a series of progressively penalized problems. This approach involves starting with \( w \) at a small value and incrementally increasing it by a factor \( \vartheta > 1 \). For each subsequent problem with a higher value of \( w \), we utilize the solution of the previous penalized problem as the initialization. This iterative process continues until \( w \) reaches a pre-set threshold value.

Upon merging the task selection strategy $\{\mathcal{A}_k\}$ into the beamforming vectors design, we note that the objective function in $\mathtt{P}_{2w}$ is in a fractional structure. In the subsequent section, we will employ quadratic and Dinkelbach's transformation techniques \cite{10121446} to convert the problem into a series of concave subproblems, thus enabling efficient solutions.
\section{Maximizing Computation Efficiency by optimizing $\mathtt{P}_{2w}$}
\label{AM}
Since the optimization variables in $\mathtt{P}_{2w}$ are strongly coupled in the objective function, it is difficult to solve $\mathtt{P}_{2w}$ directly. To this end, we optimize $\mathtt{P}_{2w}$ under the AM framework. The main idea of the AM is to alternately optimize one block of variables at a time while keeping the others fixed \cite{andresen2016convergence}. Specifically, we partition the optimization problem $\mathtt{P}_{2w}$ into seven subproblems and solve them in an AM fashion. This section will cover the optimization with respect to $
\{\boldsymbol{v}_{nk}^{\mathrm{U}}\},\{\boldsymbol{v}_{nk}^{\mathrm{D}}\},\{f_{nk},f_{k}^{\min}\},\{\theta_m^{\mathrm{U}}\},\{a_k\}$ and $\{R_k,t_k\}$. The next section will be dedicated to the optimization of $\{\theta_m^{\mathrm{D}}\}$.
\subsection{Optimizing Variables $\{\boldsymbol{v}_{nk}^{\mathrm{D}}\}$}
\label{4b}
When other variables are fixed, the subproblem for updating $\{\boldsymbol{v}_{nk}^{\mathrm{D}}\}$ is
\begin{subequations}
\label{vnk}
\begin{align}
\label{vnko}
&\min_{\substack{\left\{\!\boldsymbol
{v}_{nk}^{\mathrm{D}}\!\right\}}}  \sum_{n=1}^{N}\sum_{k=1}^{K}\mathbf{1}_{\!\left\{\boldsymbol{v}_{nk}^{\mathrm{D}} \neq \mathbf{0}\right\}} \!\!\left(\!\frac{\left(C_kt_kR_{k}f_{nk}\kappa_n\right)}{T}\!+\!\left\|\boldsymbol{v}_{n k}^{\mathrm{D}}\right\|_{2}^{2}\!\right)\!,\\
\label{vnk1}
&\text { s.t. }
\mathbf{1}_{\left\{a_k \neq 0\right\}} \text{SINR}_k^{\mathrm{D}}\left(\{{\boldsymbol{v}_{nk}^{\mathrm{D}}}\}, \{\theta_m^{\mathrm{D}}\}\right) \geq \mathbf{1}_{\left\{a_k \neq 0\right\}} \gamma_{k}^{\mathrm{D}}, \  k \in \mathcal{K},\\
\label{vnk2}
&\quad\quad\sum_{k \in \mathcal{K}}\left\|\boldsymbol{v}_{n k}^{\mathrm{D}}\right\|_{2}^{2} \leq P_{n, \max }^{\mathrm{D}},\ \  \forall n \in \mathcal{N},\\
\begin{split}
\label{vnmlatency}
&\quad\quad\mathbf{1}_{\left\{\boldsymbol{v}_{nk}^{\mathrm{D}} \neq \mathbf{0}\right\}}\!\Biggl(\!\frac{U_k -TR_k^{\text{loc}}}{R_k}\!+\!\frac{U_k -TR_k^{\text{loc}}}{f_{nk}}\!\Biggl) \leq \mathbf{1}_{\left\{\boldsymbol{v}_{nk}^{\mathrm{D}} \neq \mathbf{0}\right\}} \chi, \\
&\quad \quad \quad \quad \quad \quad \quad \quad \quad \quad \quad\quad\quad\quad\quad k \in \mathcal{K}, n \in \mathcal{N},
\end{split}\\
\label{18e}
&\quad\quad\mathbf{1}_{\left\{\boldsymbol{v}_{nk}^{\mathrm{D}} \neq \mathbf{0}\right\}} R_k \leq \mathbf{1}_{\left\{\boldsymbol{v}_{nk}^{\mathrm{D}} \neq \mathbf{0}\right\}} R_{nk}, \quad k \in \mathcal{K}, n \in \mathcal{N},\\
\label{frequencynorm1}
&\quad\quad \mathbf{1}_{\left\{\boldsymbol{v}_{nk}^{\mathrm{D}} \neq \mathbf{0}\right\}}f_k^{\min}\leq\mathbf{1}_{\left\{\boldsymbol{v}_{nk}^{\mathrm{D}} \neq \mathbf{0}\right\}}f_{nk}, \  k \in \mathcal{K}, n \in \mathcal{N},\\
\label{scac}
&\quad\quad\sum_{n=1}^{N} \sum_{k=1}^{K} \left|\|[(\boldsymbol{v}_{n k}^{\mathrm{D}})^{\mathrm{T}}\ (\boldsymbol{v}_{n k}^{\mathrm{U}})^{\mathrm{T}}]\|_2\right| \leq \beta.
\end{align}
\end{subequations}
Since an arbitrary phase rotation of vector $\boldsymbol{v}_{n k}^{\mathrm{D}}$ does not affect the SINR constraints (\ref{vnk1}), we can replace the non-convex constraints (\ref{vnk1}) with second-order cone (SOC) constraints \cite{6832894}:
\begin{equation}
\label{SOC}
\begin{aligned}
&\mathbf{1}_{\left\{a_k \neq 0\right\}}\sqrt{\sum_{l \neq k}\left|\sum_{n \in \mathcal{N}}\left(\boldsymbol{\iota}_{n k}^{\mathrm{D}}\right)^{\mathrm{H}} \boldsymbol{v}_{n l}^{\mathrm{D}}\right|^{2}+\sigma_{k}^{2}} \\
&\leq \mathbf{1}_{\left\{a_k \neq 0\right\}}\frac{1}{\sqrt{\gamma_{k}^{\mathrm{D}}}} \Re\left(\sum_{n \in \mathcal{N}}\left(\boldsymbol{\iota}_{n k}^{\mathrm{D}}\right)^{\mathrm{H}} \boldsymbol{v}_{n k}^{\mathrm{D}}\right), \quad \forall k \in \mathcal{K}.
\end{aligned}
\end{equation}

For the objective function in (\ref{vnk}), we identify that it has an indicator function which is equivalent to a weighted group $\ell_{0}$-"norm" of $\left[\left(\boldsymbol{v}_{11}^{\mathrm{D}}\right)^{\!\mathrm{T}}\!, \cdots, \left(\boldsymbol{v}_{1K}^{\mathrm{D}}\right)^{\!\mathrm{T}}\!, \cdots, \left(\boldsymbol{v}_{NK}^{\mathrm{D}}\right)^{\!\mathrm{T}}\right]^{\!\mathrm{T}}$, as given by $\sum_{n=1}^{N}\sum_{k=1}^{K} \mathbf{1}_{\left\{\boldsymbol{v}_{nk}^{\mathrm{D}} \neq \mathbf{0}\right\}}$, with weights $\!\left(C_kt_kR_{k}f_{nk}\kappa_n\!\right)\!/T\!+\!\left\|\boldsymbol{v}_{n k}^{\mathrm{D}}\right\|_{2}^{2}$ \cite{seneviratne2012l0}, and it is non-convex. As group $\ell_{1}$-norm is the tightest convex relaxation to group $\ell_{0}$-"norm", the objective function of (\ref{vnk}) can be relaxed as $\sum_{n=1}^{N} \sum_{k=1}^{K}\left\|\boldsymbol{v}_{nk}^{\mathrm{D}}\right\|_{2} \Bigl(\!\left(C_kt_kR_{k}f_{nk}\kappa_n\right)/T+\left\|\boldsymbol{v}_{n k}^{\mathrm{D}}\right\|_{2}^{2}\Bigr)$. 
Furthermore, considering the constraints (\ref{vnmlatency}), (\ref{18e}) and (\ref{frequencynorm1}), because the $\ell_{0}$-"norm" and $\ell_{1}$-norm are always non-negative, replacing the $\ell_{0}$-"norm" with the $\ell_{1}$-norm will not change the relationship of the inequality. Therefore, the constraints (\ref{vnmlatency}), (\ref{18e}) and (\ref{frequencynorm1}) can be transformed into equivalent forms as
\begin{equation}
\label{30}
\begin{aligned}
\left\|\boldsymbol{v}_{nk}^{\mathrm{D}}\right\|_{2}\Biggl(\frac{U_k-TR_k^{\text{loc}}}{R_k} + \frac{U_k -TR_k^{\text{loc}}}{f_{nk}}\Biggl) &\leq \left\|\boldsymbol{v}_{nk}^{\mathrm{D}}\right\|_{2}  \chi, \\
\quad k &\in \mathcal{K}, n \in \mathcal{N},
\end{aligned}
\end{equation}
\begin{equation}
\label{21}
\left\|\boldsymbol{v}_{nk}^{\mathrm{D}}\right\|_{2}R_k \leq \left\|\boldsymbol{v}_{nk}^{\mathrm{D}}\right\|_{2}R_{nk}, \quad k \in \mathcal{K}, n \in \mathcal{N},
\end{equation}
and
\begin{equation}
\label{frequencynorm2}
\left\|\boldsymbol{v}_{nk}^{\mathrm{D}}\right\|_{2}f_k^{\min}\leq\left\|\boldsymbol{v}_{nk}^{\mathrm{D}}\right\|_{2}f_{nk}, \  k \in \mathcal{K}, n \in \mathcal{N},
\end{equation}
respectively. 

Therefore, the transformed problem of (\ref{vnk}) is given as
\begin{equation}
\label{equa}
\begin{aligned}
\min_{\substack{\left\{\boldsymbol
{v}_{nk}^{\mathrm{D}}\right\}}} 
\quad&\sum_{n=1}^{N} \sum_{k=1}^{K}\left\|\boldsymbol{v}_{nk}^{\mathrm{D}}\right\|_{2} \!\!\left(\!\frac{\left(C_kt_kR_{k}f_{nk}\kappa_n\right)}{T}\!+\!\left\|\boldsymbol{v}_{n k}^{\mathrm{D}}\right\|_{2}^{2}\right),\\
\text { s.t. }
\quad& \text{(\ref{vnk2})},\text{(\ref{scac})},\text{(\ref{SOC})},\text{(\ref{30})},\text{(\ref{21})},\text{(\ref{frequencynorm2})}.
\end{aligned}
\end{equation}
Since (\ref{equa}) is a convex optimization problem, it can be directly solved via the interior-point method\cite{polik2010interior}. Theoretical results from \cite{1255564} show that the optimal solution to (\ref{equa}) provides a feasible solution to (\ref{vnk}).
\subsection{Optimizing Variables $\{f_{nk}\}$ and $\{f_{k}^{\min}\}$}
Regarding the optimization of $\{f_{nk}\}$ and $\{f_{k}^{\mathrm{min}}\}$, the subproblem is given as
\begin{equation}
\label{updatef}
\begin{aligned}
\min_{\substack{\{f_{nk}\}, \{f_{k}^{\mathrm{min}}\}}} \  &\sum_{n=1}^{N}\!\sum_{k=1}^{K}\mathbf{1}_{\left\{\boldsymbol{v}_{nk}^{\mathrm{D}} \neq \mathbf{0}\right\}}\!\!\left(\!\frac{t_kR_{k}}{T}f_{nk}\kappa_nC_k\!\right)\\
\text{s.t.} \ \  &(\text{\ref{frequency}}),\text{(\ref{tk1})},(\text{\ref{latencynorm}}),(\text{\ref{freqnorm}}).
\end{aligned}
\end{equation}
By multiplying both sides of the constraint (\ref{latencynorm}) by $f_{nk}$ and both sides of the constraint (\ref{tk1}) by $f_{k}^{\min}$, these constraints are transformed into linear constraints in terms of $f_{nk}$ and $f_{k}^{\min}$, respectively. Consequently, this manipulation renders the optimization problem (\ref{updatef}) as a convex optimization problem, which can be efficiently solved using the interior-point method.
\subsection{Optimizing Variables $\{\boldsymbol{v}_{nk}^{\mathrm{U}}\}$}

Considering $\mathtt{P}_{2w}$, when the remaining variables are held fixed, the subproblem for updating $\{\boldsymbol{v}_{nk}^{\mathrm{U}}\}$ is
\begin{subequations}
\label{vnku}
\begin{align}
\max_{\substack{\{\boldsymbol{v}_{nk}^{\mathrm{U}}\}}} \quad 
&\frac{1}{w}\sum_k\sum_n\mathbf{1}_{\left\{\boldsymbol{v}_{nk}^{\mathrm{D}} \neq \mathbf{0}\right\}}\log( R_{nk}\left(\boldsymbol{v}_{nk}^{\mathrm{U}}\right)- R_k),\\
\text { s.t. } \quad 
&\sum_{n=1}^{N} \sum_{k=1}^{K} \left|\|[(\boldsymbol{v}_{n k}^{\mathrm{D}})^{\mathrm{T}}\ (\boldsymbol{v}_{n k}^{\mathrm{U}})^{\mathrm{T}}]\|_2\right| \leq \beta.
\end{align}
\end{subequations}
Since $R_{nk}\left(\boldsymbol{v}_{nk}^{\mathrm{U}}\right)$ has a fractional structure, we employ a quadratic transform to reconfigure it as
\begin{subequations}
\label{vnkqt}
\begin{align}
\max_{\substack{\{\boldsymbol{v}_{nk}^{\mathrm{U}}\},\boldsymbol{s}}}&\frac{1}{w}\!\sum_k\!\sum_n\!\mathbf{1}_{\left\{\boldsymbol{v}_{nk}^{\mathrm{D}}\neq \mathbf{0}\right\}}\!\log(\log_2\!\left(Q_{nk}(\boldsymbol{v}_{nk}^{\mathrm{U}})\right)\!- \!R_k),\\
\text { s.t. } \ \  
&\sum_{n=1}^{N} \sum_{k=1}^{K} \left|\|[(\boldsymbol{v}_{n k}^{\mathrm{D}})^{\mathrm{T}}\ (\boldsymbol{v}_{n k}^{\mathrm{U}})^{\mathrm{T}}]\|_2\right| \leq \beta,
\end{align}
\end{subequations}
where
\begin{equation}
\begin{aligned}
&Q_{nk}(\boldsymbol{v}_{nk}^{\mathrm{U}})={1+2\text{Re}(s_{nk}p_{k}^{\mathrm{U}}(\boldsymbol{\iota}_{nk}^{\mathrm{U}})^{\mathrm{H}}\boldsymbol{v}_{nk}^{\mathrm{U}})}\\
\!-&{s_{nk}\!\!\left(\!\sum_{l \neq k} p_{l}^{\mathrm{U}}\!\!\left(\boldsymbol{v}_{n k}^{\mathrm{U}}\right)^{\mathrm{H}}\!\!
\boldsymbol{\iota}_{nl}^{\mathrm{U}}(\boldsymbol{\iota}_{nl}^{\mathrm{U}})^{\mathrm{H}}\boldsymbol{v}_{n k}^{\mathrm{U}}
\!+\!\sigma_{n}^{2}(\boldsymbol{v}_{n k}^{\mathrm{U}})^{\mathrm{H}}\boldsymbol{v}_{n k}^{\mathrm{U}}\!\!\right)\!\!s_{nk}^{\mathrm{H}}}
\end{aligned}
\end{equation}
and $\boldsymbol{s}=\{s_{nk}\}$ is the set of auxiliary variables introduced through the quadratic transformation. Then iteratively solving (\ref{vnkqt}) can obtain a stationary point of (\ref{vnku}). In particular, when $\{\boldsymbol{v}_{nk}^{\mathrm{U}}\}$ are held fixed, the optimal $s_{nk}$ is given by
\begin{equation}  
\label{ss}
\!\!\!s_{nk}^{\star}\!=\!\frac{p_{k}^{\mathrm{U}}\left(\boldsymbol{v}_{nk}^{\mathrm{U}}\right)^{\mathrm{H}}\!
\boldsymbol{\iota}_{nk}^{\mathrm{U}}}{\left(\!\sum_{l \neq k} p_{l}^{\mathrm{U}}\!\left(\boldsymbol{v}_{n k}^{\mathrm{U}}\right)^{\mathrm{H}} 
\!\boldsymbol{\iota}_{nl}^{\mathrm{U}}(\boldsymbol{\iota}_{nl}^{\mathrm{U}})^{\mathrm{H}}\boldsymbol{v}_{n k}^{\mathrm{U}}\!
\!+\!\sigma_{n}^{2}(\boldsymbol{v}_{n k}^{\mathrm{U}})^{\mathrm{H}}\boldsymbol{v}_{n k}^{\mathrm{U}}\!\right)}.
\end{equation}
On the other hand, when $\boldsymbol{s}$ is fixed, (\ref{vnkqt}) becomes a convex problem with respect to $\boldsymbol{v}_{nk}^{\mathrm{U}}$. This is because $Q_{nk}(\boldsymbol{v}_{nk}^{\mathrm{U}})$ is a concave function with respect to $\boldsymbol{v}_{nk}^{\mathrm{U}}$, and the composition of $\log$ and $Q_{nk}(\boldsymbol{v}_{nk}^{\mathrm{U}})$ preserves concavity. Since the subproblem for solving $\boldsymbol{v}_{nk}^{\mathrm{U}}$ is a convex optimization problem, it can be efficiently solved using the interior-point method.
\subsection{Optimizing Variables $\{\theta_m^{\mathrm{U}}\}$}
When it comes to the uplink RIS reflecting coefficients design, the subproblem for updating $\{\theta_m^{\mathrm{U}}\}$ is given by
\begin{equation}
\label{thetau}
\max _{\substack{\{\theta_m^{\mathrm{U}}\}}} \quad 
\frac{1}{w}    \sum_k\sum_n\mathbf{1}_{\left\{\boldsymbol{v}_{nk}^{\mathrm{D}} \neq \mathbf{0}\right\}}\log( R_{nk}\left(\{\theta_m^{\mathrm{U}}\}\right)- R_k),\\
\end{equation}
where we substitute (\ref{UU}) in the objective function. Since (\ref{thetau}) is an unconstrained optimization problem with a differentiable objective function, it can be readily solved by the gradient descent method. To be specific, the update of  $\theta_m^{\mathrm{U}}$ at the $i^{t h}$ iteration is given by
\begin{equation}
\label{gradientdescent}
\begin{aligned}
&\theta_m^{\mathrm{U}}\!\left(i+1\right)=\theta_m^{\mathrm{U}}(i)\\
&+\mathcal{I}(i)\Bigl(\frac{1}{w}\sum\limits_{n\in \mathcal{N}}\sum\limits_{k\in \mathcal{K}}\mathbf{1}_{\left\{\boldsymbol{v}_{nk}^{\mathrm{D}} \neq \mathbf{0}\right\}}\frac{\nabla_{\theta_m^{\mathrm{U}}}R_{nk}\left(\theta_m^{\mathrm{U}}(i)\right)}{R_{nk}\left(\{\theta_m^{\mathrm{U}}\}\right)- R_k}\Bigr),
\end{aligned}
\end{equation}
where $\mathcal{I}(i)$ is the step size chosen by the Armijo rule to guarantee convergence \cite{li2022phase} and $\nabla_{\theta_m^{\mathrm{U}}}R_{nk}\left(\theta_m^{\mathrm{U}}(i)\right)$ is the gradient of $R_{nk}\left(\theta_m^{\mathrm{U}}\right)$ at point $\theta_m^{\mathrm{U}}(i)$ as given by (\ref{nabla}). By updating $\{\theta_m^{\mathrm{U}}\}$ based on (\ref{gradientdescent}), it is guaranteed to converge to a stationary point of the problem (\ref{thetau}).
\begin{figure*}
\vspace*{-5mm}
\begin{gather}
\label{nabla}
\begin{split}
\nabla_{\theta_m^{\mathrm{U}}}R_{nk}\left(\theta_m^{\mathrm{U}}(i)\right)=\operatorname{Re}\Bigg(\frac{\psi_m(i)}{\left(1+\gamma_{n k}^{\mathrm{U}}\right)\ln2}\bigg(
\frac{2p_{k}^{\mathrm{U}}\left(\boldsymbol{v}_{nk}^{\mathrm{U}}\right)^{\mathrm{H}}\!
\boldsymbol{\iota}_{nk}^{\mathrm{U}}\left(\boldsymbol{K}_{nkk}\right)_{m,m}
}
{\sum_{l \neq k} p_{l}^{\mathrm{U}}\left|\left(\boldsymbol{v}_{n k}^{\mathrm{U}}\right)^{\mathrm{H}} 
\boldsymbol{\iota}_{nl}^{\mathrm{U}}
\right|^{2}\!+\!\sigma_{n}^{2}\left\|\boldsymbol{v}_{n k}^{\mathrm{U}}\right\|_{2}^{2}}\\
-\frac{p_{k}^{\mathrm{U}}\left|\left(\boldsymbol{v}_{nk}^{\mathrm{U}}\right)^{\mathrm{H}}\!\boldsymbol{\iota}_{n k}^{\mathrm{U}}\right|^{2}\sum_{l \neq k} 2p_{l}^{\mathrm{U}}\left(\boldsymbol{v}_{n k}^{\mathrm{U}}\right)^{\mathrm{H}}\boldsymbol{\iota}_{n l}^{\mathrm{U}}\left(\boldsymbol{K}_{nkl}\right)_{m,m}}
{\left(\sum_{l \neq k} p_{l}^{\mathrm{U}}\left|\left(\boldsymbol{v}_{n k}^{\mathrm{U}}\right)^{\mathrm{H}} \boldsymbol{\iota}_{nl}^{\mathrm{U}}\right|^{2}\!+\!\sigma_{n}^{2}\left\|\boldsymbol{v}_{n k}^{\mathrm{U}}\right\|_{2}^{2}\right)^2}\bigg)\Bigg)
\end{split}
\end{gather}\\
where $\boldsymbol{K}_{nkl}=\left(\boldsymbol{G}_{n}^{\mathrm{U}}\right)^{*}\left(\boldsymbol{v}_{nk}^{\mathrm{U}}\right)^{*}\left(\boldsymbol{h}_{\mathrm{r}, l}^{\mathrm{U}}\right)^{\mathrm{T}}$ and $\psi_m(i)=je^{j\theta_m^{\mathrm{U}}(i)}\left(\beta_{\min}+2^{-\alpha}(1-\beta_{\min})\Big(1+\sin\big(\theta_m^{\mathrm{U}}(i)-\phi\big)\Big)^{\alpha}\right)+\\
    \alpha 2^{-\alpha} e^{j\theta_m^{\mathrm{U}}(i)} (1-\beta_{\min}) \Big(1+\sin(\theta_m^{\mathrm{U}}(i)-\phi)\Big)^{\alpha-1} \cos(\theta_m^{\mathrm{U}}(i)-\phi).$
\hrule  
\end{figure*}
\subsection{Optimizing Variables $\{a_k\}$}
In terms of variable $\{a_k\}$, when other variables in $\mathtt{P}_{2w}$ are fixed, the subproblem for updating $\{a_k\}$ is
\begin{subequations}
\label{subproblema}
\begin{align}
\begin{split}
\max _{\substack{\{a_k\}}} \  &\frac{\sum_{k=1}^{K}t_kR_k/T+\sum_{k=1}^{K}R_{k}^{\mathrm{loc}}(a_k)}{\sum_{n=1}^{N}\sum_{k=1}^{K} \mathbf{1}_{\left\{\boldsymbol{v}_{nk}^{\mathrm{D}} \neq \mathbf{0}\right\}} \frac{\left(C_kt_kR_{k}f_{nk}\kappa_n\right)}{T}}\\
&+\frac{1}{w}\sum_k\sum_n\mathbf{1}_{\left\{\boldsymbol{v}_{nk}^{\mathrm{D}} \neq \mathbf{0}\right\}}\log( R_{nk}\left(\{a_k\}\right)- R_k),
\end{split}\\
\label{19a}
\text { s.t. } \  &a_{k} \in[0,1], \quad \forall k \in \mathcal{K},\\
\label{19g}
    &\mathbf{1}_{\left\{a_k \neq 0\right\}} \text{SINR}_k^{\mathrm{D}}\left(\{{\boldsymbol{v}_{nk}^{\mathrm{D}}}\}, \{\theta_m^{\mathrm{D}}\}\right) \geq \mathbf{1}_{\left\{a_k \neq 0\right\}} \gamma_{k}^{\mathrm{D}}, \  k \in \mathcal{K},\\
    \begin{split}
\label{19d}
&\mathbf{1}_{\left\{\boldsymbol{v}_{nk}^{\mathrm{D}} \neq \mathbf{0}\right\}}\!\Biggl(\!\frac{U_k-TR_k^{\text{loc}}}{R_k}\!+\!\frac{U_k-TR_k^{\text{loc}}}{f_{nk}}\!\Biggl)\leq \!\mathbf{1}_{\left\{\boldsymbol{v}_{nk}^{\mathrm{D}} \neq \mathbf{0}\right\}} \chi,\\
&\quad\quad\quad\quad\quad\quad\quad\quad\quad\quad\quad\quad\quad\quad\quad\quad k \in \mathcal{K}, n \in \mathcal{N}.
\end{split}
\end{align}
\end{subequations}
By identifying that (\ref{19g}) is a weight $\ell_{0}$-"norm" as introduced in \ref{4b}, we equivalently transform it as
\begin{equation}
\label{trans19g}
\left\|a_k\right\|_{2} \text{SINR}_k^{\mathrm{D}}\left(\{{\boldsymbol{v}_{nk}^{\mathrm{D}}}\}, \{\theta_m^{\mathrm{D}}\}\right) \geq \left\|a_k\right\|_{2}\gamma_{k}^{\mathrm{D}}, \quad k \in \mathcal{K}.
\end{equation}
Next, as $\{a_k\}$ in $R_{nk}(\{a_k\})$ also possesses a fractional structure, akin to our treatment of $\{\boldsymbol{v}_{nk}^{\mathrm{U}}\}$, we also apply quadratic transformation:
\begin{equation}
\label{aq}
\begin{aligned}
\max_{\substack{\{a_k\}, \boldsymbol{o}}} \ 
&\frac{\sum_{k=1}^{K}t_kR_k/T+\sum_{k=1}^{K}R_{k}^{\mathrm{loc}}(a_k)}{\sum_{n=1}^{N}\sum_{k=1}^{K} \mathbf{1}_{\left\{\boldsymbol{v}_{nk}^{\mathrm{D}} \neq \mathbf{0}\right\}} \frac{\left(C_kt_kR_{k}f_{nk}\kappa_n\right)}{T}}\\
&+\frac{1}{w}\sum_k\sum_n\mathbf{1}_{\left\{\boldsymbol{v}_{nk}^{\mathrm{D}} \neq \mathbf{0}\right\}}\log(D_{nk}(\{a_k\})- R_k), \\
\text{s.t.} \ &\text{(\ref{19a})}, \text{(\ref{19d})},\text{(\ref{trans19g})},
\end{aligned}
\end{equation}
where $\boldsymbol{o}$ refers to the collection of $\{o_{nk}\}$ and $D_{nk}(\{a_k\})$ is given as
\begin{equation}
\begin{aligned}
D_{nk}&(\{a_k\})\!=\!\log_2\Biggl(1+2o_{nk}\sqrt{a_{k} P_{k}^{c}}\left|\left(\boldsymbol{v}_{nk}^{\mathrm{U}}\right)^{\mathrm{H}}\!
\boldsymbol{\iota}_{nk}^{\mathrm{U}}\right|\\
&-\!o_{nk}^2\biggl(\sum_{l \neq k} a_{l} P_{l}^{c}\left|\left(\boldsymbol{v}_{n k}^{\mathrm{U}}\right)^{\mathrm{H}} \boldsymbol{\iota}_{nl}^{\mathrm{U}}\right|^2\!\!+\!\sigma_{n}^{2}\|\boldsymbol{v}_{n k}^{\mathrm{U}}\|^2\biggl)\!\Biggl).
\end{aligned}
\end{equation}
Similar to (\ref{vnkqt}), when $\{a_k\}$ is held fixed, the optimal $o_{nk}$ can be directly given by
\begin{equation}
\label{fop}
o_{nk}^{\star}=\frac{a_{k} P_{k}^{c}\left|\left(\boldsymbol{v}_{nk}^{\mathrm{U}}\right)^{\mathrm{H}}\!\boldsymbol{\iota}_{nk}^{\mathrm{U}}\right|^2}{\sum_{l \neq k} a_{l} P_{l}^{c}\left|\left(\boldsymbol{v}_{n k}^{\mathrm{U}}\right)^{\mathrm{H}} \boldsymbol{\iota}_{nl}^{\mathrm{U}}\right|^2+\sigma_{n}^{2}\left|\boldsymbol{v}_{n k}^{\mathrm{U}}\right|^2}.
\end{equation} When $\boldsymbol{o}$ is fixed, (\ref{aq}) is a convex problem with respect to $a_k$, and therefore, the optimal solution can be efficiently found using the standard numerical method.
\subsection{Optimizing Variables $\{R_k\}$ and $\{t_k\}$}
Regarding the optimization of $\{R_k\}$ and $\{t_k\}$, the subproblem is given as
\begin{subequations} 
\begin{align}
\begin{split}
\max _{\substack{\{R_k\},\{t_k\}}} \  &\frac{\sum_{k=1}^{K}t_kR_k/T+\sum_{k=1}^{K}R_{k}^{\mathrm{loc}}}{\sum_{n=1}^{N}\!\sum_{k=1}^{K}\!\!\mathbf{1}_{\left\{\boldsymbol{v}_{nk}^{\mathrm{D}} \neq \mathbf{0}\right\}}\!\!\left(\!\frac{t_kR_{k}}{T}f_{nk}\kappa_nC_k\!\right)}\!\\
&+\frac{1}{w}\underbrace{\sum_k\sum_n\mathbf{1}_{\left\{\boldsymbol{v}_{nk}^{\mathrm{D}} \neq \mathbf{0}\right\}}\left(\log( R_{nk}- R_k)\right)}_{:=e\left(\{R_k\}\right)},
\end{split}\\
\label{19tk}
\text{s.t.}\quad & \ t_k+(C_kt_kR_k)/f_k^{\min} \leq T, \quad \forall k \in \mathcal{K},\\
&(\text{\ref{19d}})\nonumber.
\end{align}
\end{subequations}
Then applying Dinkelbach's transformation to separate the numerator and denominator, we have
\begin{equation}
\label{Rk}
\begin{aligned}
\max_{\substack{\{R_k\},\{t_k\},\\
z}} &\ \ 
\frac{1}{w}e\left(\{R_k\}\right)+{\sum_{k=1}^{K}\left(\frac{t_kR_k}{T}+R_{k}^{\mathrm{loc}}\right)}\\
-z&\left({\sum_{n=1}^{N}\sum_{k=1}^{K} \mathbf{1}_{\left\{\boldsymbol{v}_{nk}^{\mathrm{D}} \neq \mathbf{0}\right\}} \left(\!\frac{t_kR_k}{T}f_{nk}\kappa_nC_k\!\right)}\right),\\
\text{s.t.} \ \  &(\text{\ref{19d}}),(\text{\ref{19tk}}),
\end{aligned}
\end{equation}
where $z$ is the auxiliary variable.
Similarly, when $\{R_k\}$ and
$\{t_k\}$ are fixed, the optimal $z$ is given by the iterative update rule of the Dinklebach transform as 
\begin{equation}
\label{zs}
z^{\star}=\frac{\sum_{k=1}^{K}t_kR_k/T+\sum_{k=1}^{K}R_{k}^{\mathrm{loc}}}{\sum_{n=1}^{N}\!\sum_{k=1}^{K}\!\!\mathbf{1}_{\left\{\boldsymbol{v}_{nk}^{\mathrm{D}} \neq \mathbf{0}\right\}}\!\!\left(\!\frac{t_kR_k}{T}f_{nk}\kappa_nC_k\!\right)}.
\end{equation}

With other variables fixed, the subproblem for updating $\{R_k\}$ is given as
\begin{equation}
\label{solvefnk}
\begin{aligned}
\max_{\substack{\{R_k\}}} \ \ 
&\frac{1}{w}e\left(\{R_k\}\right)+{\sum_{k=1}^{K}\left(\frac{t_kR_k}{T}+R_{k}^{\mathrm{loc}}\right)}\\
&-z\left({\sum_{n=1}^{N}\sum_{k=1}^{K} \mathbf{1}_{\left\{\boldsymbol{v}_{nk}^{\mathrm{D}} \neq \mathbf{0}\right\}} \left(\!\frac{t_kR_k}{T}f_{nk}\kappa_nC_k\!\right)}\right),\\
\text{s.t.} \ \  &(\text{\ref{19d}}),(\text{\ref{19tk}}).
\end{aligned}
\end{equation}
The constraints (\ref{19d}) can be linearized by multiplying both sides of the constraint by $R_k$. This transformation converts the optimization problem (\ref{solvefnk}) into a convex one, which can then be solved using standard numerical methods.

Finally, regarding $\{t_k\}$, with other variables are fixed, the subproblem for updating  $\{t_k\}$ is given as
\begin{equation}
\label{Rkonly}
\begin{aligned}
\max_{\substack{\{t_k\}}} \ \ 
&\sum_{k=1}^{K}\left(\frac{t_kR_k}{T}+R_{k}^{\mathrm{loc}}\right)\\
&-z\left({\sum_{k=1}^{K}\sum_{n=1}^{N}\mathbf{1}_{\left\{\boldsymbol{v}_{nk}^{\mathrm{D}} \neq \mathbf{0}\right\}} \left(\!\frac{t_kR_k}{T}f_{nk}\kappa_nC_k\!\right)}\right),\\
\text{s.t.} \ \  &(\text{\ref{19tk}}).
\end{aligned}
\end{equation}
It is important to note that the optimization problem (\ref{Rkonly}) is a linear programming problem. Therefore, it can be solved using standard numerical methods.

\section{Designing $\{\theta_m^{\mathrm{D}}\}$ Beyond a Random Feasible Point in $\mathtt{P}_{2w}$}
\label{F}
Unlike other variables in $\mathtt{P}_{2w}$, $\{\theta_m^{\mathrm{D}}\}$ does not appear in the objective function of $\mathtt{P}_{2w}$. This absence of dependence often leads to a misconception that any $\{\theta_m^{\mathrm{D}}\}$ satisfying constraint (\ref{b}) would be a valid solution, as has been done in previous works \cite{8811733,8741198,9457078}. However, it is known that an optimized RIS together with properly designed beamforming vector $\{\boldsymbol{v}_{nk}^{\mathrm{D}}\}$ at the APs can help to construct a high-quality channel link, thus reduce the downlink transmission power at APs while satisfying the SINR constraints (\ref{b}). This intuition contradicts to simply finding a feasible solution for (\ref{b}).

Recognizing that the variables $\{\theta_m^{\mathrm{D}}\}$ work cooperatively with $\{\boldsymbol{v}_{nk}^{\mathrm{D}}\}$ to reduce the downlink transmission power, we resort to the optimization problem (\ref{vnk}) that updates $\{\boldsymbol{v}_{nk}^{\mathrm{D}}\}$. More insight can be revealed by noticing that $\{\theta_m^{\mathrm{D}}\}$ are hidden in $\boldsymbol{\Theta}^{\mathrm{D}}$ and is coupled with $\boldsymbol{v}_{nk}^{\mathrm{D}}$ in the SINR constraints (\ref{SOC}) (an equivalent SOC form of (\ref{b})). We can potentially obtain a better solution of (\ref{vnk}) by adjusting  $\{\theta_m^{\mathrm{D}}\}$ such that the active constraints in (\ref{SOC}) are turned into inactive constraints\footnote{If the equality holds, the constraint is active (tight), otherwise inactive.}\cite{heath2018scientific}. 

Motivated by the fact that the difference between the right-hand-side and the left-hand-side of equation (\ref{SOC}) is bigger than zero if the constraint is inactive and equals zero if it is active, we compute the minimum differences between the right-hand-side and the left-hand-side of the equation (\ref{SOC}) among the $K$ users. Then, we tune the $\{\theta_m^{\mathrm{D}}\}$ such that the minimum difference is maximized, giving the following maximin optimization problem:
\begin{subequations}
\label{activecon}
\begin{align}
\begin{split}
    \max _{\{\theta_m^{\mathrm{D}}\}}\min_{\substack{k \in \mathcal{K},\\a_k \neq 0}} \  &\frac{1}{\sqrt{\gamma_{k}^{\mathrm{D}}}} \Re\left(\sum_{n \in \mathcal{N}}\left(\boldsymbol{h}_{\mathrm{d}, n k}^{\mathrm{D}}+\left(\boldsymbol{G}_{n}^{\mathrm{D}}\right)^{\mathrm{H}}\boldsymbol{\Theta}^{\mathrm{D}} \boldsymbol{h}_{\mathrm{r}, k}^{\mathrm{D}}\right)^{\mathrm{H}} \boldsymbol{v}_{n k}^{\mathrm{D}}\right) \\
    & \!\!\!\!\!\!\!\!\!\!\!\!\!\!\!\!\!\!\!\!\!\!\!\!\!\!\!\!\! - \sqrt{\sum_{l \neq k}\left|\sum_{n \in \mathcal{N}}\left(\boldsymbol{h}_{\mathrm{d}, n k}^{\mathrm{D}}+\left(\boldsymbol{G}_{n}^{\mathrm{D}}\right)^{\mathrm{H}}\boldsymbol{\Theta}^{\mathrm{D}} \boldsymbol{h}_{\mathrm{r}, k}^{\mathrm{D}}\right)^{\mathrm{H}} \boldsymbol{v}_{n l}^{\mathrm{D}}\right|^{2}+\sigma_{k}^{2}},
    \end{split}\\\label{34b}
    \text{s.t.} \quad \quad & \boldsymbol{\Theta}^{\mathrm{D}}=\operatorname{diag}\left(\rho(\theta_{1}^{\mathrm{D}})e^{j\theta_{1}^{\mathrm{D}}}, \ldots, \rho(\theta_{M}^{\mathrm{D}})e^{j\theta_{M}^{\mathrm{D}}}\right), \nonumber \\
    &\quad \forall m \in \mathcal{M},
\end{align}
\end{subequations}
where we expressed $\boldsymbol{\iota}_{n k}^{\mathrm{D}}$ as $\boldsymbol{h}_{\mathrm{d}, n k}^{\mathrm{D}}+\left(\boldsymbol{G}_{n}^{\mathrm{D}}\right)^{\mathrm{H}}\boldsymbol{\Theta}^{\mathrm{D}} \boldsymbol{h}_{\mathrm{r}, k}^{\mathrm{D}}$ according to (\ref{downlinkchannel}).

By introducing a slack variable $y$, the above optimization problem can be recast as \cite{boyd2004convex}
\begin{subequations}
\label{slack}
\begin{align}
&\max _{\{\theta_m^{\mathrm{D}}\}, y} \quad y,\\
\begin{split}
\label{8.31}
&\text { s.t. }
\frac{1}{\sqrt{\gamma_{k}^{\mathrm{D}}}} \Re\left(\sum_{n \in \mathcal{N}}\left(\boldsymbol{h}_{\mathrm{d}, n k}^{\mathrm{D}}+\left(\boldsymbol{G}_{n}^{\mathrm{D}}\right)^{\mathrm{H}}\boldsymbol{\Theta}^{\mathrm{D}} \boldsymbol{h}_{\mathrm{r}, k}^{\mathrm{D}}\right)^{\mathrm{H}} \boldsymbol{v}_{n k}^{\mathrm{D}}\right)\\
\quad& - \sqrt{\sum_{l \neq k}\left|\sum_{n \in \mathcal{N}}\!\left(\boldsymbol{h}_{\mathrm{d}, n k}^{\mathrm{D}}+\left(\boldsymbol{G}_{n}^{\mathrm{D}}\right)^{\mathrm{H}}\boldsymbol{\Theta}^{\mathrm{D}} \boldsymbol{h}_{\mathrm{r}, k}^{\mathrm{D}}\right)^{\mathrm{H}}\!\!\boldsymbol{v}_{n l}^{\mathrm{D}}\right|^{2}\!\!+\!\sigma_{k}^{2}} \! \geq \! y, \\
&\quad \quad \quad \quad \quad \bigcap_{\substack{k \in \mathcal{K},a_k \neq 0}} \{k\},
\end{split}\\
\label{slack-prac}
\quad & \boldsymbol{\Theta}^{\mathrm{D}}=\operatorname{diag}\left(\rho(\theta_{1}^{\mathrm{D}})e^{j\theta_{1}^{\mathrm{D}}}, \ldots, \rho(\theta_{M}^{\mathrm{D}})e^{j\theta_{M}^{\mathrm{D}}}\right), \quad \forall m \in \mathcal{M}.
\end{align}
\end{subequations}
Furthermore, to deal with the non-convexity introduced by the practical phase shift model in (\ref{slack-prac}), we treat $\boldsymbol{\Theta}^{\mathrm{D}}$ as an auxiliary variable and resort to the penalty-based method by adding an equality constraint-related penalty term to the objective function of
(\ref{slack}), yielding the following optimization problem
\begin{subequations}
\label{penactiveslack}
\begin{align}
\begin{split}
&\max_{\{\theta_m^{\mathrm{D}}\},\boldsymbol{\Theta}^{\mathrm{D}},y} y \\
& \quad \quad \quad -\mu \left\lVert\boldsymbol{\Theta}^{\mathrm{D}} - \operatorname{diag}\left(\rho(\theta_{1}^{\mathrm{D}})e^{j\theta_{1}^{\mathrm{D}}}, \ldots, \rho(\theta_{M}^{\mathrm{D}})e^{j\theta_{M}^{\mathrm{D}}}\right) \right\rVert_{\text{F}}^2,
\end{split}
\\
&\text { s.t. } \quad \quad \text{(\ref{8.31})},\nonumber
\\
\label{z} & \ \ \quad\quad\quad\left(\boldsymbol{\Theta}^{\mathrm{D}}\right)_{i,j} = 0, \quad \forall i \neq j,
\end{align}
\end{subequations}
where $\mu>0$ is a penalty parameter for controlling the constraint violation of (\ref{slack-prac})\cite{8410057} and constraints (\ref{z}) is to restrict the non-diagonal entries of $\boldsymbol{\Theta}^{\mathrm{D}}$ are to be zero. It is observed that $\{\theta_m^{\mathrm{D}}\}$ can be updated with fixed $\boldsymbol{\Theta}^{\mathrm{D}}$ and $y$. On the other hand, $\boldsymbol{\Theta}^{\mathrm{D}}$ and $y$ can be updated in parallel when $\{\theta_m^{\mathrm{D}}\}$ are fixed. This motivates us to apply the AM method to solve (\ref{penactiveslack}) efficiently. 

For any given $\boldsymbol{\Theta}^{\mathrm{D}}$ and $y$, the subproblem for optimizing $\{\theta_m^{\mathrm{D}}\}$ is 
\begin{equation}
\label{algo-theta}
\max _{\{\theta_m^{\mathrm{D}}\}} \ -\mu \left\lVert\boldsymbol{\Theta}^{\mathrm{D}} - \operatorname{diag}\left(\rho(\theta_{1}^{\mathrm{D}})e^{j\theta_{1}^{\mathrm{D}}}, \ldots, \rho(\theta_{M}^{\mathrm{D}})e^{j\theta_{M}^{\mathrm{D}}}\right) \right\rVert_{\text{F}}^2,
\end{equation}
which is an unconstrained optimization problem with a differentiable objective function, and it can be readily solved by the gradient descent method. On the other hand, for any given $\{\theta_m^{\mathrm{D}}\}$, the subproblem for optimizing $\boldsymbol{\Theta}^{\mathrm{D}}$ and $y$ is 
\begin{equation}
\label{t}
\begin{aligned}
\max _{\boldsymbol{\Theta}^{\mathrm{D}}, y} \quad &y -\mu \left\lVert\boldsymbol{\Theta}^{\mathrm{D}} - \operatorname{diag}\left(\rho(\theta_{1}^{\mathrm{D}})e^{j\theta_{1}^{\mathrm{D}}}, \ldots, \rho(\theta_{M}^{\mathrm{D}})e^{j\theta_{M}^{\mathrm{D}}}\right) \right\rVert_{\text{F}}^2,\\
\text{s.t.} \quad & (\text{\ref{8.31})}, (\text{\ref{z}}).
\end{aligned}
\end{equation}
Noticing that (\ref{8.31}) are convex constraints since they are SOC constraints that composite with affine function. This makes the optimization problem (\ref{t}) a convex problem and can be directly solved via the interior-point method.
The entire procedure for solving (\ref{activecon}) is summarized in Algorithm 2, where we employ an increasing penalty strategy to successively enforce the constraint of (\ref{34b}). Initially, $\mu$ is set to be a sufficiently small value. By gradually increasing the value of $\mu$ by a factor $\varrho>1$, we can maximize the original objective function and obtain a solution that satisfies the equality constraints in (\ref{34b}) within a predefined accuracy. 
\RestyleAlgo{ruled}
\begin{algorithm}[t]
    \SetKwInOut{Input}{Input}
    \SetKwInOut{Output}{Output}
\caption{Penalty-based Algorithm for Solving (\ref{activecon})}\label{thetadl}
\Input{$\{\boldsymbol{v}_{nk}^{\mathrm{D}}\}$ and set $\mu>0$.}
\While{$\left\lVert\boldsymbol{\Theta}^{\mathrm{D}} - \operatorname{diag}\left(\rho(\theta_{1}^{\mathrm{D}})e^{j\theta_{1}^{\mathrm{D}}}, \ldots, \rho(\theta_{M}^{\mathrm{D}})e^{j\theta_{M}^{\mathrm{D}}}\right) \right\rVert_{\mathrm{F}}^2 > \epsilon_1$}{
\While{The fractional increase of the objective value of (\ref{penactiveslack}) is above $\epsilon_2$}{Update $\theta_m^{\mathrm{D}}$ by solving (\ref{algo-theta}) with fixed  $\boldsymbol{\Theta}^{\mathrm{D}}$ and $y$.\\
Update $\boldsymbol{\Theta}^{\mathrm{D}}$ and $y$ by solving (\ref{t}) with fixed  $\theta_m^{\mathrm{D}}$.}
Update the penalty coefficient $\mu \leftarrow \varrho \mu $.}
\end{algorithm}
\section{Finding Initial Feasible Points and Complexity Analysis}
Sections IV and V describe how to solve $\mathtt{P}_{2w}$ for a fixed $w$. To enforce the log barrier, the penalty weight $w$ should be increased progressively after $\mathtt{P}_{2w}$ is solved with a particular $w$. This leads to Algorithm 2 for solving the overall problem $\mathtt{P}_{2}$, with the last line corresponding to the increase of $w$. Once we solve $\mathtt{P}_{2w}$ for a particular $w$, the solution would act as the initialization for the next $\mathtt{P}_{2w}$ with a larger $w$. However, a natural question arises regarding the initialization for $\mathtt{P}_{2w}$ for the first selected $w$. Since a feasible point for $\mathtt{P}_{2}$ is also a feasible point for $\mathtt{P}_{2w}$ with any $w$, the following section will find a feasible point of $\mathtt{P}_{2}$.

\subsection{Finding Initial Feasible Points}
Given the complicated constraints of $\mathtt{P}_2$, it is not easy to find a feasible point. Inspired by \cite{lipp2016variations,8474356}, we propose an effective heuristic method to address this challenge. The idea is to address a modified feasibility problem. This involves allowing violations of the selected constraints in the original problem while penalizing them within the objective function of the modified feasibility problem. Subsequently, these violations gradually decrease through the optimization process and fade away, ultimately leading to a feasible solution for the original problem. Particularly, let us consider the modification of the feasibility problem of $\mathtt{P}_{2}$ as in (\ref{yyy}), where we penalized the downlink SINR constraints (\ref{b}). 
\begin{figure*}
\begin{equation}
\label{yyy}
\begin{aligned}
    \max _{\substack{\{\boldsymbol{v}_{nk}^{\mathrm{U}}\},\{\boldsymbol{v}_{nk}^{\mathrm{D}}\}, \{R_k\},\{f_{nk}\},\\ \{\theta_m^{\mathrm{U}}\},\{\theta_m^{\mathrm{D}}\}, \{a_k\},\{f_{k}^{\min}\},\{t_k\}}}  \  &\sum_{ k \in \mathcal{K}}\left\|a_k\right\|_{2} \left[\frac{1}{\sqrt{\gamma_{k}^{\mathrm{D}}}} \Re\left(\sum_{n \in \mathcal{N}}\left(\boldsymbol{h}_{\mathrm{d}, n k}^{\mathrm{D}}+\left(\boldsymbol{G}_{n}^{\mathrm{D}}\right)^{\mathrm{H}}\boldsymbol{\Theta}^{\mathrm{D}} \boldsymbol{h}_{\mathrm{r}, k}^{\mathrm{D}}\right)^{\mathrm{H}} \boldsymbol{v}_{n k}^{\mathrm{D}}\right) \right.\\
    &  \quad \quad - \left.\sqrt{\sum_{l \neq k}\left|\sum_{n \in \mathcal{N}}\left(\boldsymbol{h}_{\mathrm{d}, n k}^{\mathrm{D}}+\left(\boldsymbol{G}_{n}^{\mathrm{D}}\right)^{\mathrm{H}}\boldsymbol{\Theta}^{\mathrm{D}} \boldsymbol{h}_{\mathrm{r}, k}^{\mathrm{D}}\right)^{\mathrm{H}} \boldsymbol{v}_{n l}^{\mathrm{D}}\right|^{2}+\sigma_{k}^{2}},\right]^{+}\\
    \text{s.t.} \quad & (\text{\ref{a}}),(\text{\ref{d}}),(\text{\ref{UU}}),(\text{\ref{tk1}}),(\text{\ref{frequency}}),(\text{\ref{latencynorm}}),(\text{\ref{normmaximin}}),(\text{\ref{freqnorm}}), (\text{\ref{pn}})
\end{aligned}
\end{equation}
\hrule  
\end{figure*}
Maximizing the objective function in (\ref{yyy}) enhances downlink SINR and reduces constraint violations in (\ref{b}). If the objective function value of (\ref{yyy}) surpasses zero, it signifies the violations disappear, and the solution of (\ref{yyy}) is a feasible solution for $\mathtt{P}_{2}$.

Solving (\ref{yyy}) involves a straightforward process. Initially, we generate random values for $\{\theta_m^{\mathrm{U}}\}$, $\{\theta_m^{\mathrm{D}}\}$, and $\{a_k\}$. Following this, we generate $\{\boldsymbol{v}_{nk}^{\mathrm{D}}\}$ and adjust their scaling to satisfy the conditions in (\ref{d}). By utilizing $\{\boldsymbol{v}_{nk}^{\mathrm{D}}\}$, $\{\theta_m^{\mathrm{U}}\}$, $\{\theta_m^{\mathrm{D}}\}$ and $\{a_k\}$, $\{\boldsymbol{v}_{nk}^{\mathrm{U}}\}$, $\{R_k\}$, $\{f_{nk}\}$, $\{f_{k}^{\min}\}$ and $\{t_k\}$ are determined to adhere to the equalities stipulated in equations (\ref{frequency}), (\ref{tk1}), (\ref{latencynorm}), (\ref{normmaximin}), (\ref{freqnorm}) and (\ref{pn}).
With this initial set of optimization variables established, we can initiate an iterative AM procedure to tackle the optimization problem (\ref{yyy}). To begin with, when the other variables are held fixed, each individual variable in (\ref{yyy}) exhibits convex behavior in the objective function, which is a summation of SOCs. Moreover, concerning the nonconvex constraints, the strategies for transforming them into convex constraints have been given in Sections \ref{4b} through \ref{F}. As a result, all subproblems of the variables can be transformed into convex forms and subsequently solved. This enables the utilization of the AM procedure to solve (\ref{yyy}) efficiently, and the details of these steps are omitted for brevity.

\subsection{Complexity Analysis of the Proposed Algorithm}
The computational complexity of Algorithm 2 primarily comes from solving problems (\ref{vnku}) and (\ref{vnk}), which aim to design \(\{\boldsymbol{v}_{nk}^{\mathrm{U}}\}\) and \(\{\boldsymbol{v}_{nk}^{\mathrm{D}}\}\), respectively, as well as the downlink RIS reflecting coefficients \(\{\theta_m^{\mathrm{D}}\}\), and problem (\ref{subproblema}) for optimizing the power partition parameters \(\{a_k\}\). The complexity associated with resolving (\ref{vnku}) and (\ref{vnk}) is estimated to be \(\mathcal{O}((NKL)^{3.5})\), based on the complexities of solving convex problems using the interior point method \cite{5447068}. Similarly, the task of optimizing \(\{a_k\}\) is estimated to have a computational complexity of \(\mathcal{O}(K^{3.5})\), following the same computational approach. The complexity of employing Algorithm 1 via the penalty-based method is approximated as \(\mathcal{O}(M^{3.5})\). If we denote the number of iterations for solving problems (\ref{vnku}), (\ref{vnk}), and (\ref{subproblema}) as \(I_1\), \(I_2\), and \(I_3\), respectively, and those for Algorithm 1 as \(I_4\), then the total computational complexity is given as \(\mathcal{O}(I((I_1 + I_2)(NKL)^{3.5} + I_3K^{3.5} + I_4M^{3.5}))\), where \(I\) represents the total number of iterations within Algorithm 2.
\RestyleAlgo{ruled}
\begin{algorithm}[t]
    \SetKwInOut{Input}{Input}
    \SetKwInOut{Output}{Output}
\caption{Algorithm for Solving $\mathtt{P}_{2}$}\label{p2}
\Input{Initial $\{\boldsymbol{v}_{nk}^{\mathrm{D}}\}$, $\{a_k\}$,$\{R_k\}$, $\{t_k\}$, $\{\theta_m^{\mathrm{U}}\}$, $\{\theta_m^{\mathrm{D}}\}$, $\{f_{nk}\}$,$
\{f_{k}^{\min}\}$ and $w>0$.}
\While{Stopping criterion is not satisfied}{
\While{Stopping criterion is not satisfied}{
Update $\boldsymbol{s}$ using equation (\ref{ss}).\\
Update $\{\boldsymbol{v}_{nk}^{\mathrm{U}}\}$ by solving (\ref{vnkqt}) with fixed $\boldsymbol{s}$.\\}
\While{Stopping criterion is not satisfied}{
Update $\boldsymbol{o}$ using equation (\ref{fop}).\\
Update $\{a_k\}$ by solving (\ref{aq}) with fixed $\boldsymbol{o}$.
}
\While{Stopping criterion is not satisfied}{
Update $\{\theta_m^{\mathrm{U}}\}$ using equation (\ref{gradientdescent}).
}
Update $\{\boldsymbol{v}_{nk}^{\mathrm{D}}\}$ by solving (\ref{equa}).\\
Update $\{f_{nk}\}$ and $\{f_{k}^{\min}\}$ by solving (\ref{updatef}).\\
\While{Stopping criterion is not satisfied}{
Update $z$ using equation (\ref{zs}).\\
Update $\{R_k\}$ by solving (\ref{Rkonly}) with fixed $z$ and $\{t_k\}$.\\
Update $\{t_k\}$ by solving (\ref{solvefnk}) with fixed $z$ and $\{R_k\}$.
}
Update $\{\theta_m^{\mathrm{D}}\}$ by Algorithm 1.\\
Update the penalty coefficient $w \leftarrow \vartheta w $.
}
\end{algorithm}

\section{Simulation Results}
In this section, we present simulation results to verify the effectiveness of our proposed algorithm. Under a three-dimensional Cartesian coordinate system, we consider a system with 10 APs and 20 users uniformly and randomly distributed in a square region of $200 \mathrm{~m} \times 200 \mathrm{~m}$. An RIS with 20 reflecting elements is located at the 3-dimensional coordinate $(100,0,15)$. In addition, the APs are with height $30 \mathrm{~m}$, while the users are with height $1 \mathrm{~m}$.

Rician fading channel model is considered for all channels to account for both the line-of-sight (LoS) and non-LoS (NLoS) components \cite{tse2005fundamentals}. For example, the AP-RIS channel can be expressed as $\boldsymbol{G}_{n}^{x}=\sqrt{L_{\mathrm{AR}}(d)}\left(\sqrt{\frac{\kappa_{\mathrm{AR}}}{1+\kappa_{\mathrm{AR}}}} \boldsymbol{G}_{n}^{\mathrm{LoS}}+\sqrt{\frac{1}{1+\kappa_{\mathrm{AR}}}} \boldsymbol{G}_{n}^{\mathrm{NLoS}}\right)$, where $\kappa_{\mathrm{AR}}$ is the Rician factor representing the ratio of power between the LoS path and the scattered paths, $\boldsymbol{G}_{n}^{\mathrm{LoS}}$ is the LoS component modeled as the product of the unit spatial signature of the AP-RIS link \cite{tse2005fundamentals}, $\boldsymbol{G}_{n}^{\text {NLoS}}$ is the Rayleigh fading components with entries distributed as $\mathcal{C} \mathcal{N}(0,1)$, $L_{\mathrm{AR}}(d)$ is the distance-dependent path loss of the AP-RIS channel, and $x \in\{\mathrm{U}, \mathrm{D}\}$. We consider the following distance-dependent path loss model $L_{\mathrm{AR}}(d)=10^{0.3}E_{0}\left(\frac{d}{d_{0}}\right)^{-\alpha_{\mathrm{AR}}}$, where $E_{0}$ is the constant path loss at the reference distance $d_{0}=1 \mathrm{~m}, d$ is the Euclidean distance between the transceivers, $\alpha_{\mathrm{AR}}$ is the path loss exponent, and $10^{0.3}$ accounts for a $3~\mathrm{dBi}$ gain at each element of the RIS since it reflects signal only in its front half-space \cite{8941080}. Since the RIS can be practically deployed in LoS with the AP, we set $\alpha_{\mathrm{AR}}=2$ and $\kappa_{\mathrm{AR}}=30 \mathrm{~dB}$ \cite{8811733}. In addition, other channels are similarly generated with $\alpha_{\mathrm{AU}}=3.67$ and $\kappa_{\mathrm{AU}}=0$ (i.e., Rayleigh fading to account for rich scattering) for the AP-user channel, $\alpha_{\mathrm{RU}}=2.5$ and $\kappa_{\mathrm{RU}}=3$ for the RIS-user channel. We consider a system with a bandwidth $10 \mathrm{~MHz}$ and $E_{0}=-30 \mathrm{~dB}$. The effective noise power for the APs and users are $\sigma_{n}^{2}=-60~\mathrm{dBm}$ and $\sigma_{k}^{2}=-50~\mathrm{dBm}$, respectively. Unless specified otherwise, other parameters are set as follows: $P_{n, \max }^{\mathrm{D}}=1 \mathrm{~W}, P_{k}^c=0.5 \mathrm{~W}$, $C_k=200 \mathrm{~cycles/bit}$, $\kappa_k = \kappa_n = 10^{-25}$, $f_n=1.2\times10^9 \mathrm{~bits/s}$, $T=0.5\mathrm{~s}$, $\chi=0.4\mathrm{~s}$ and $U_k=350\mathrm{~Kb}$ \cite{9380744,9133107}. For the proposed algorithm, we set $\mu = 10^{-3}$, $\varrho = 1.003$, $\epsilon_1 = 0.1$, $\epsilon_2 = 0.01$, and the convergence threshold in terms of the relative increment in the objective value as $10^{-3}$. The simulation results are obtained by averaging over 100 simulation trials.

We compare the proposed algorithm with the following benchmarks.

\begin{itemize}
    \item Without-RIS: Without the deployment of an RIS, the equivalent channels in (\ref{uplinkchannel}) and (\ref{downlinkchannel}) contain only the direct link, i.e., $\boldsymbol{h}_{\mathrm{r}, k}^{\mathrm{U}}=\boldsymbol{h}_{\mathrm{r}, k}^{\mathrm{D}}=\mathbf{0}, \ \forall k$. 
    \item Without cooperative transmission (denoted as Without-CT): In this case, we assume that each user is served by only one AP. For each user, the AP with the best channel condition is selected.
    \item Without RIS and cooperative transmission (denoted as Without-RIS-CT): In this case, we assume that each user is served by only one AP without the assistance of an RIS. 
    \item Proposed algorithm based on solving Feasibility Problem of $\{\theta_m^{\mathrm{D}}\}$ (denoted as AM-FP): Instead of optimizing $\{\theta_m^{\mathrm{D}}\}$ by our algorithm, in this case, we only find a feasible solution for $\{\theta_m^{\mathrm{D}}\}$ that satisfied (\ref{b}). This benchmark is designed to reveal the necessity of the proposed criterion for optimizing $\{\theta_m^{\mathrm{D}}\}$ in (\ref{activecon}).
    \item Proposed algorithm without $l_{12}$-norm (denoted as AM-Exhaustive Search (AM-ES)): In this case, we do not eliminate the variable $\{\mathcal{A}_k\}$ by exploiting the group sparsity structure of the beamforming vectors but search over all the possibilities of $\{\mathcal{A}_k\}$. This benchmark is designed to reveal the efficiency of our proposed algorithm.
\end{itemize}



\begin{figure}[t]
     \centering
     \begin{subfigure}[b]{0.44\textwidth}
             \centering   
     \includegraphics[width=\textwidth]{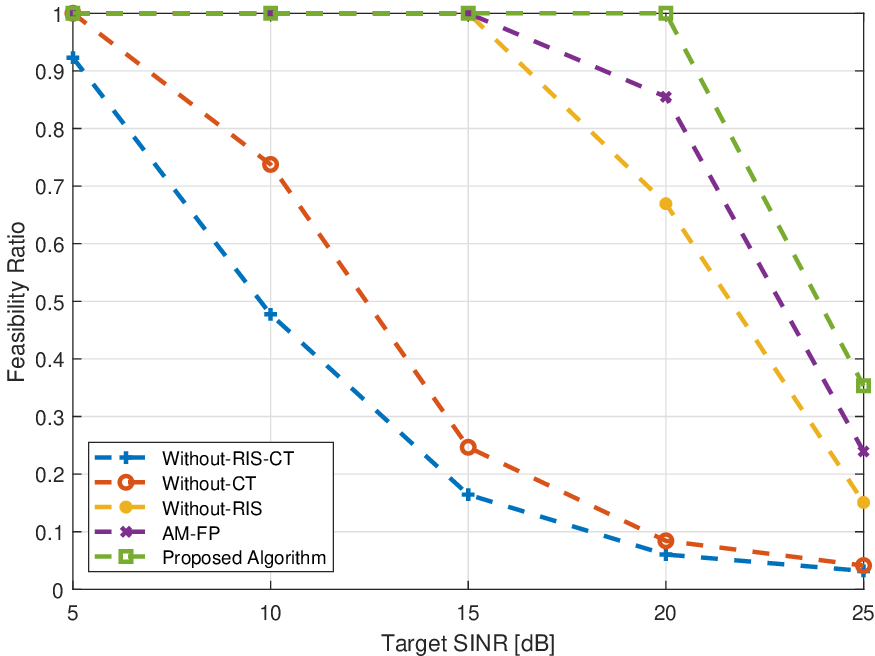}
        \caption{Feasibility ratio.}
        \label{fig:table}
     \end{subfigure}
     \hfil
     \begin{subfigure}[b]{0.44\textwidth}
      \centering
\centerline{\includegraphics[width=\textwidth]{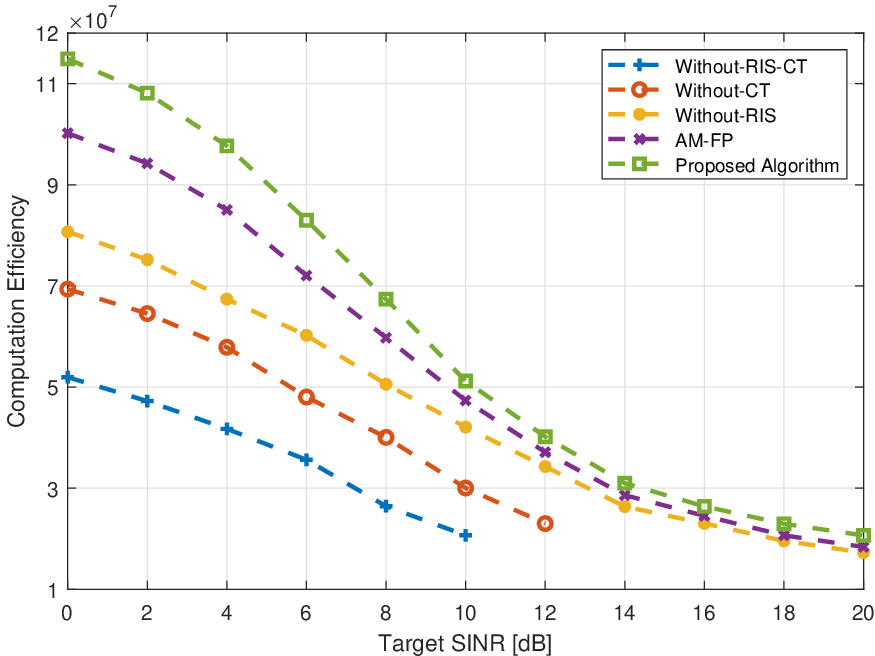}}
\caption{Computation efficiency.}
\label{EE}
     \end{subfigure}
\caption{Performance comparison versus target SINR $\gamma_{k}^{\mathrm{D}}$.}
        \label{fig}
\end{figure}

We first study the relationship between the feasibility ratio\cite{8741198} of the problem (\ref{originorigin}) and the target SINR $\gamma_{k}^{\mathrm{D}}$ with different benchmarks. The feasibility ratio of the problem (\ref{originorigin}) is defined as $\frac{\text {number of feasible simulations for (\ref{originorigin})}}{\text { the total number of simulations }}.$ As the target SINR requirements become more stringent, i.e., larger values of $\gamma_{k}^{\mathrm{D}}$, the feasibility ratio of the problem (\ref{originorigin}) is expected to decline. As shown in Fig. \ref{fig:table}, we observe that Without-RIS-CT and Without-CT almost fail to maintain feasibility in those settings with a target SINR higher than $10 \mathrm{~dB}$, while Without-RIS, AM-FP, and proposed algorithm can still maintain feasibility with a high probability. Comparing cooperative transmission and RIS, we observed that including cooperative transmission leads to a better feasibility ratio than including an RIS. This can be seen from the fact that Without-RIS can provide more than a $10 \mathrm{~dB}$ gain compared with Without-RIS-CT, while the improvement of Without-CT from Without-RIS-CT is less than $3 \mathrm{~dB}$. Obviously, including both RIS and cooperative transmission (proposed algorithm and AM-FP) has the highest feasibility ratio, with the proposed algorithm having an even higher feasibility ratio than the AM-FP. This shows the importance of optimizing $\{\theta_m^{\mathrm{D}}\}$, rather than simply finding a feasible solution for it.

\begin{figure}[ht]
\centering
\centerline{\includegraphics[scale=0.55]{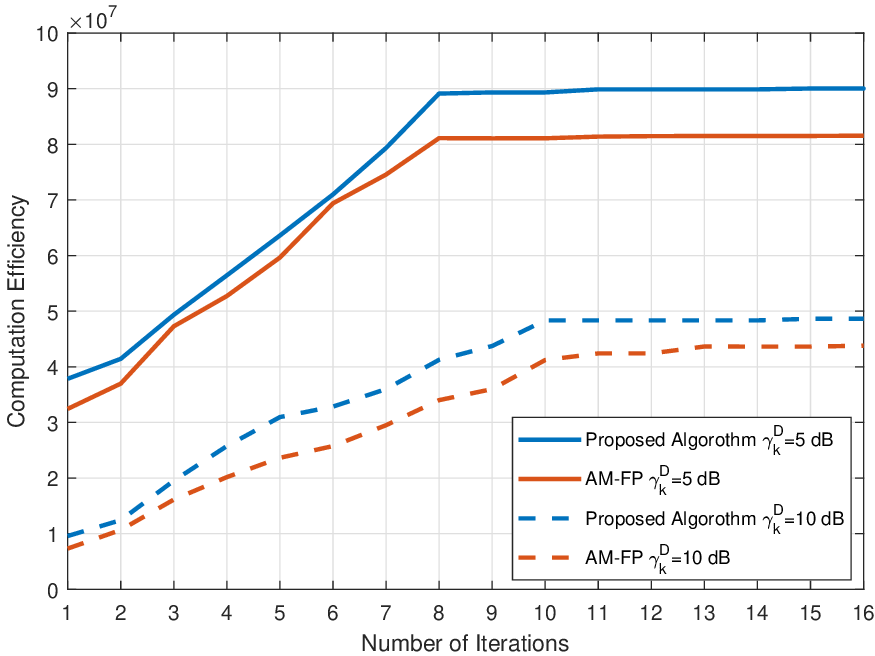}}
\caption{Convergence behaviors of both proposed algorithm and AM-FP.}
\label{conv}
\end{figure}

Next, the superiority of our proposed algorithm in terms of computation efficiency is shown in Fig. \ref{EE}. Under a wide range of SINR requirements, it is observed that the system without cooperative transmission achieves much worse computation efficiency than other schemes. The Without-RIS scheme has a $40\%$ performance improvement compared with the Without-RIS-CT scheme. A similar amount of gain can also be observed when comparing AM-FP and Without-CT. The feasibility ratio and computation efficiency results demonstrate that cooperative transmission in wireless communication systems can dramatically boost the SINR and, in turn, increase the overall computation efficiency. For the schemes with an RIS, the proposed algorithm provides at least a $14\%$ improvement in computation efficiency over AM-FP. Such a performance gain comes from the fact that the AM-FP only finds a feasible solution of $\{\theta_m^{\mathrm{D}}\}$, while the proposed algorithm optimizes $\{\theta_m^{\mathrm{D}}\}$ to improve the computation efficiency further. To make this more explicit, Fig. \ref{conv} shows the achieved computation efficiency in the first 16 iterations of the proposed algorithm and AM-FP under a specific channel realization. It is observed that the proposed algorithm outperforms AM-FP in every iteration. This shows the proposed algorithm's effectiveness in inactivating the active constraints that impede an increase in the objective value, as discussed in Section \ref{F}.

\begin{figure*}[t]
     \centering
     \begin{subfigure}[b]{0.44\textwidth}
         \centering
         \includegraphics[width=\textwidth]{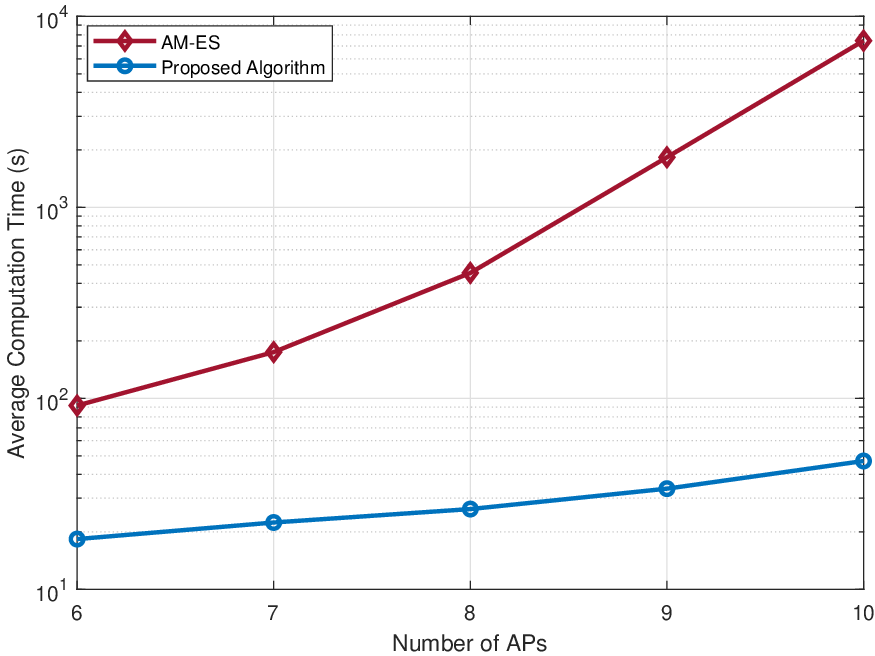}
         \caption{}
         \label{ctap}
     \end{subfigure}
     \hfil
     \begin{subfigure}[b]{0.44\textwidth}
         \centering
         \includegraphics[width=\textwidth]{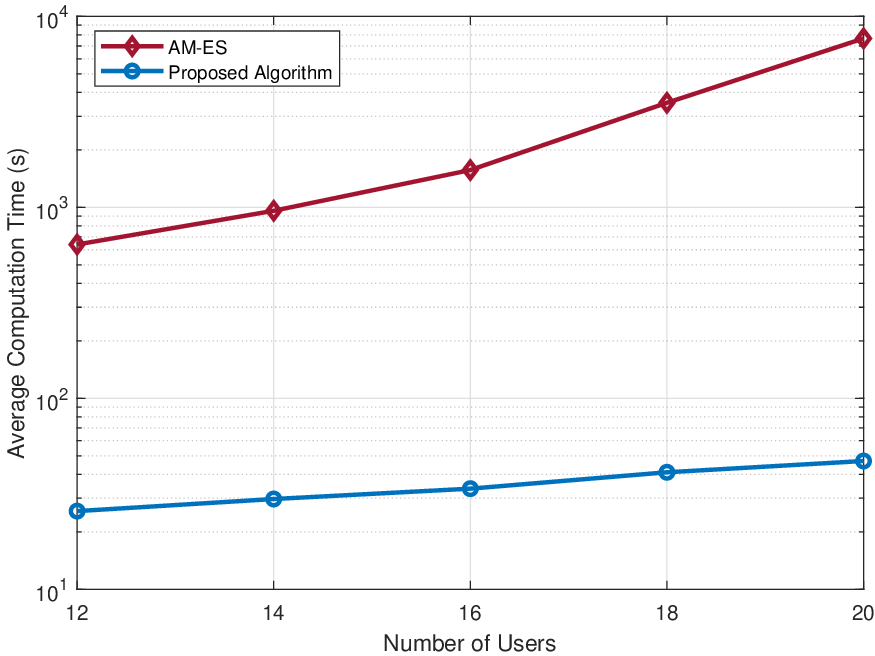}
         \caption{}
         \label{ctue}
     \end{subfigure}
     \begin{subfigure}[b]{0.44\textwidth}
         \centering
         \includegraphics[width=\textwidth]{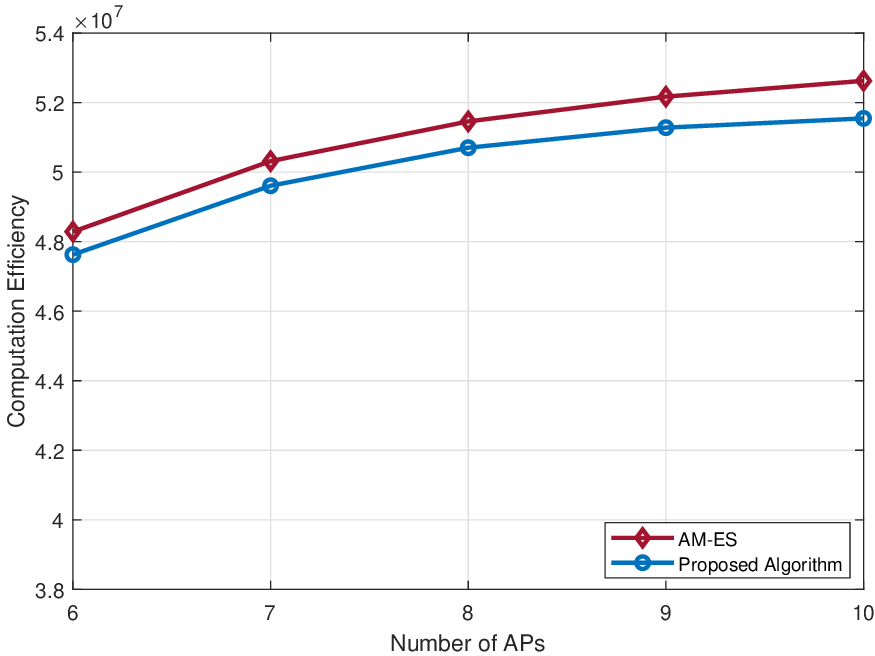}
         \caption{}
         \label{eeap}
     \end{subfigure}
     \hfil
     \begin{subfigure}[b]{0.44\textwidth}
         \centering
         \includegraphics[width=\textwidth]{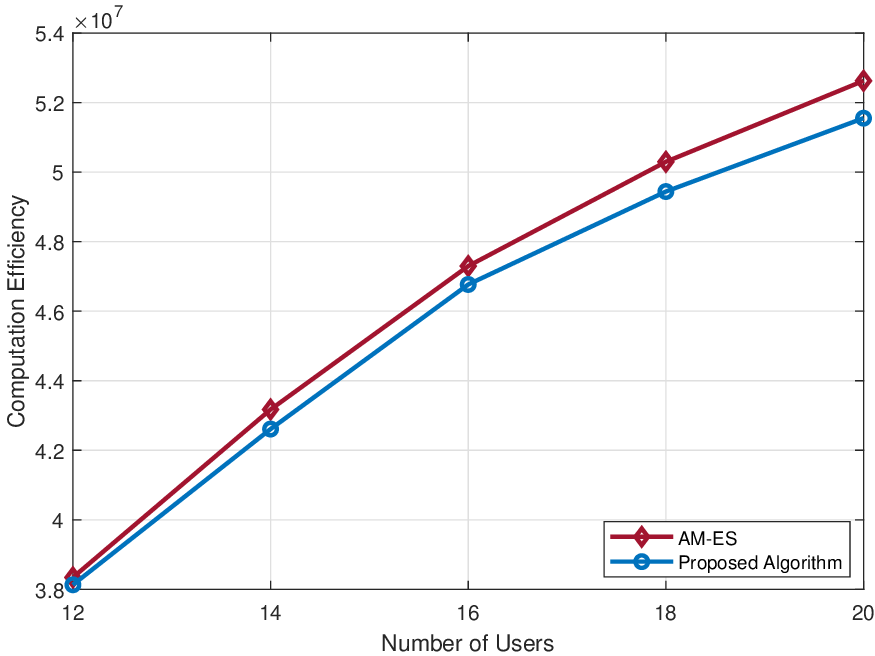}
         \caption{}
         \label{eeue}
     \end{subfigure}
        \caption{Performance comparison with the AM-ES with target SINR equals 10 dB. The number of users in (a) and (c) is 20, while the number of APs in (b) and (d) is 10.}
        \label{fig:three graphs}
\end{figure*}

To further show the complexity advantage of the proposed algorithm, we compare it with AM-ES. As shown in Fig. \ref{ctap} and Fig. \ref{ctue}, with the number of APs and users increase, the proposed algorithm saves the computation time to a large extent compared with the AM-ES (e.g., more than 10 times differences with 8 APs and 12 users, in Fig. \ref{ctap} and Fig. \ref{ctue}, respectively), and the advantage becomes more prominent as $K$ or $N$ increases. On the other hand, Fig. \ref{eeap} and Fig. \ref{eeue} show that the proposed algorithm achieves almost the same computation efficiency as the AM-ES, and the performance gap is less than $5\%$. It is also noted that in Fig. \ref{eeap} and \ref{eeue}, the total computation efficiency increases as the number of users and APs increases, but the increase diminishes. If we divide the total computation efficiency by the number of users and APs, the average computation efficiency will decrease as the number of APs and users increases. The diminishing returns on average computation efficiency suggest that the number of APs and users in the proposed system need not be arbitrarily large to obtain a favorable computation efficiency. A similar phenomenon on the system's efficiency is also exhibited in \cite{9764640,8097026}.

\begin{figure*}[t]
     \centering
     \begin{subfigure}[b]{0.48\textwidth}
         \centering
         \includegraphics[width=\textwidth]{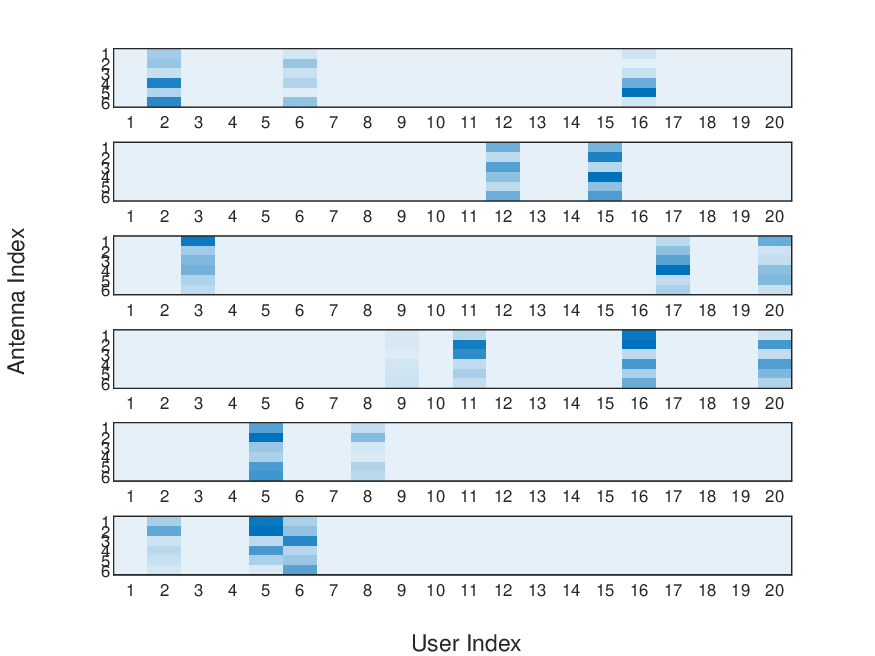}
         \caption{{Proposed Method}}
         \label{pmheat}
     \end{subfigure}
     \begin{subfigure}[b]{0.48\textwidth}
         \centering
         \includegraphics[width=\textwidth]{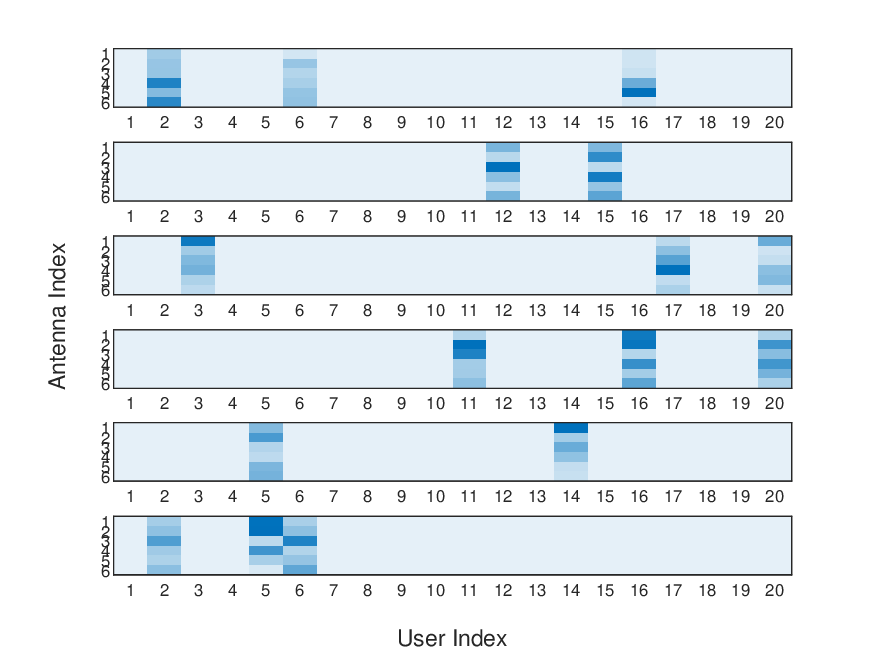}
         \caption{{AM-ES}}
         \label{amesheat}
     \end{subfigure}
        \caption{{Comparison of a heatmap of downlink beamformers with the AM-ES. The number of antennas at APs and the number of users are 6 and 20, respectively.}}
        \label{fig:heatmap}
\end{figure*}

In Figure \ref{fig:heatmap}, we present a comparison of the sparsity structure of the downlink beamformer obtained by the proposed method and the AM-ES algorithm. Each subfigure represents the downlink beamforming vectors of a specific AP, where the user index is on the horizontal axis and the antenna index is on the vertical axis. The heatmap is generated by taking the absolute value of the beamforming vectors in a particular channel realization, and the color intensity shows the magnitude of the absolute value. For brevity, we only illustrate the results for 6 APs. It can be observed that both methods yield a sparse structure in the downlink beamforming vectors, which aligns with the intuition described in Section \ref{3}. Furthermore, the solution of AM-ES has a more sparse structure than our proposed method since it reflects the optimal sparsity structure by enumerating the associations of APs and users, while our proposed method induces the sparsity structure by the group norm. The above observation explains the performance gap between our proposed method and AM-ES.

\begin{figure}[ht]
\centering
\centerline{\includegraphics[scale=0.55]{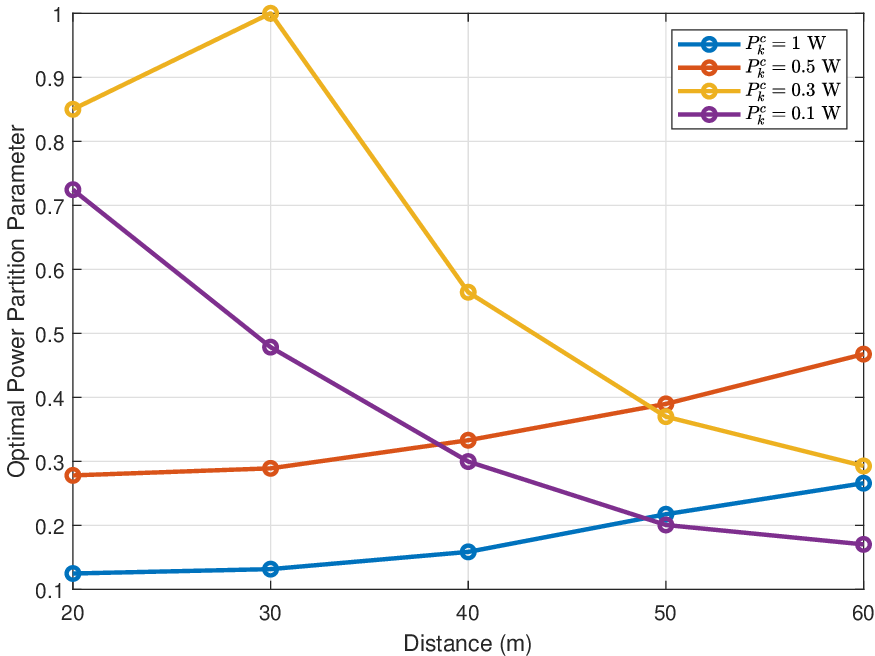}}
\caption{Optimal power partition parameter versus distance.}
\label{distance}
\end{figure}

To visualize the division of computation load between offloading and local computation, we investigate how the optimal power partition parameter varies with the distance between the user and RIS in Fig. \ref{distance}. In our scenario, we consider two users and two APs, with one user located at a fixed distance of 20 m from the RIS and the other from 20 to 60 m away. For the user with high power budgets ($P_k^c = 1$ or $0.5~\mathrm{W}$), we observe that the optimal strategy is to allocate more power to computation offloading as the distance increases. This is because computation offloading can provide a higher computation efficiency than local computation, and as the distance increases, the user needs to spend more power on data offloading.
In contrast, for a user with a low power budget ($P_k^c = 0.3~\mathrm{W}$), the optimal power partition parameter first increases as the distance increases. However, as it reaches the maximum value of $1$, it starts to decline as the distance increases. The rise in the optimal power partition parameter at the first segment is due to the same reason as users with a high power budget. However, as the distance increases further, a limited power budget cannot support high-speed data transmission in computation offloading, and computation offloading becomes less computation efficient than local computing. Thus, we observe a decline in the optimal power partition parameter at the second segment. As for the case of $P_k^c = 0.1~\mathrm{W}$, the optimal power partition parameter declines as the distance increases. The simulation results demonstrate that the optimal power allocation strategy for maximizing the computation efficiency depends on the user's power budget and the distance between the RIS and the user. In general, the optimal power partition parameter first increases as the distance increases before reaching its maximum value of $1$ and then decreases as the distance increases. The optimal power partition parameter curve will shift to the right if the power budget is abundant, while the curve will shift to the left if the budget is small.

\begin{figure}[ht]
\centering
\centerline{\includegraphics[scale=0.65]{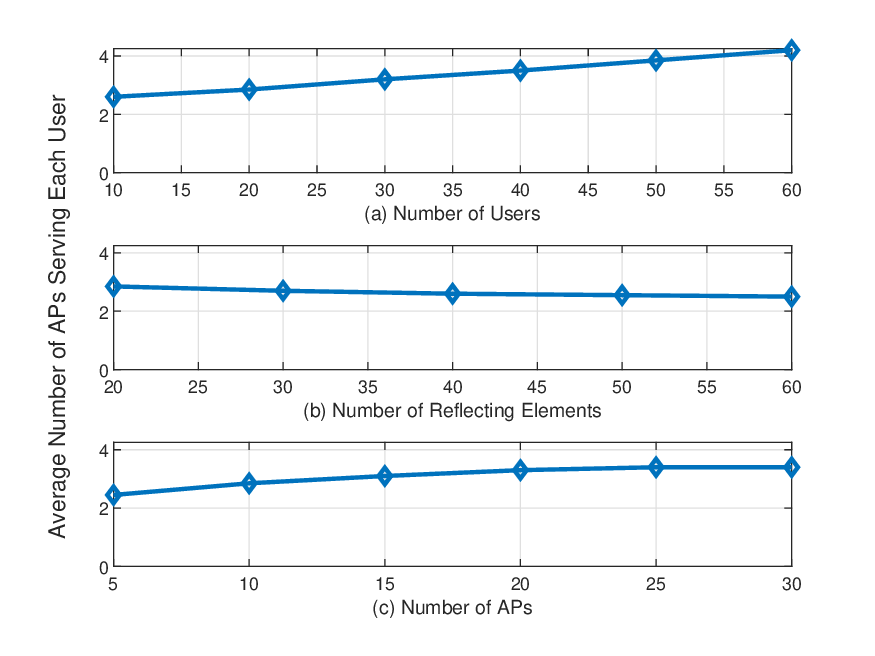}}
\caption{Average number of APs serving each user under different numbers of APs, users, and reflecting elements, with 10 APs in (a) and (b), 20 users in (b) and (c), and 20 reflecting elements in (a) and (c).}
\label{KMN}
\end{figure}

In Fig. \ref{KMN}, we show how the number of APs ($N$), users ($K$), and RIS's reflecting elements ($M$) would affect the average number of APs serving each user in user association. In Fig. \ref{KMN}(a), it is evident that as the number of users increases, there is an upward trend in the average number of APs serving each user. This can be attributed to the increased downlink interference caused by more users. Therefore, increasing the number of APs per user can reduce downlink power consumption and enhance the overall system computation efficiency. In Fig. \ref{KMN}(b), we can observe that the increasing number of reflecting elements reduces the average number of APs serving each user. This is because the increased elements of RIS improved the channel condition, allowing the system to satisfy the user's downlink QoS with fewer associated APs. In Fig. \ref{KMN}(c), as the number of APs increases, the average number of APs serving each user increases from 2.45 to 3.4 initially, then levels off after reaching 25 APs. This phenomenon occurs because having more APs per user reduces the downlink's power consumption at first. However, as the number of APs per user keeps growing, this power-saving effect becomes less pronounced, and it is outweighed by the power consumption resulting from the additional computation load of extra APs. As a result, the proposed algorithm would not further increase the association of APs and users.

\begin{figure}[ht]
\centering
\centerline{\includegraphics[scale=0.55]{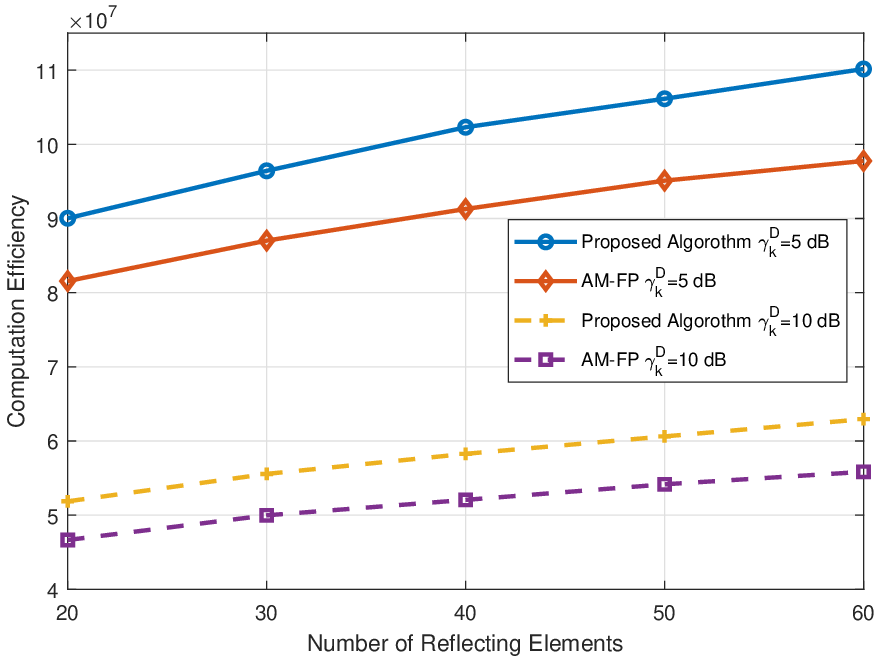}}
\caption{Computation efficiency versus the number of reflecting elements.}
\label{number}
\end{figure}

Finally, in Fig. \ref{number}, we compare the computation efficiency versus the number of reflecting elements $M$. As $M$ increases, the computation efficiency increases moderately in both the proposed algorithm and AM-FP. It is worth mentioning that the performance gaps between the proposed algorithm and AM-FP are getting more prominent as the number of reflecting elements increases, indicating that the proposed algorithm is especially appealing when the RIS is equipped with a large number of elements.

\section{Conclusion}
\label{con}
In this paper, an RIS-aided mobile edge computing system with a cooperative transmission framework was proposed. Specifically, the computation efficiency was maximized via the joint design of user association, receive/downlink beamforming vectors, power partition parameters, and uplink/downlink phase-shift matrices. For an efficient algorithm design, the alternating maximization framework was employed. To handle the discrete user association variables, group $l_{12}$-norm was adopted to enforce group sparsity and merge user association into the receive/downlink beamforming vectors. Furthermore, although the objective function does not explicitly depend on the downlink phase-shift matrix, we leverage their hidden relationship to convert it into an explicit form for optimization. This approach enables us to fully exploit the potential of the RIS rather than simply finding a feasible solution of the downlink phase-shift matrix that could be more optimal. Numerical results demonstrated that cooperative transmission and RIS could significantly improve the computation efficiency and feasibility ratio. At the same time, the proposed approach for optimizing downlink phase shifts can further promote this improvement.

\bibliographystyle{IEEEtran}
\bibliography{IEEEabrv,my}

\begin{thebibliography}{10}
\providecommand{\url}[1]{#1}
\csname url@samestyle\endcsname
\providecommand{\newblock}{\relax}
\providecommand{\bibinfo}[2]{#2}
\providecommand{\BIBentrySTDinterwordspacing}{\spaceskip=0pt\relax}
\providecommand{\BIBentryALTinterwordstretchfactor}{4}
\providecommand{\BIBentryALTinterwordspacing}{\spaceskip=\fontdimen2\font plus
\BIBentryALTinterwordstretchfactor\fontdimen3\font minus \fontdimen4\font\relax}
\providecommand{\BIBforeignlanguage}[2]{{%
\expandafter\ifx\csname l@#1\endcsname\relax
\typeout{** WARNING: IEEEtran.bst: No hyphenation pattern has been}%
\typeout{** loaded for the language `#1'. Using the pattern for}%
\typeout{** the default language instead.}%
\else
\language=\csname l@#1\endcsname
\fi
#2}}
\providecommand{\BIBdecl}{\relax}
\BIBdecl

\bibitem{8016573}
Y.~Mao, C.~You, J.~Zhang, K.~Huang, and K.~B. Letaief, ``A survey on mobile edge computing: The communication perspective,'' \emph{IEEE Commun. Surveys Tuts.}, vol.~19, no.~4, pp. 2322--2358, Fourth Quart. 2017.

\bibitem{9439524}
Y.~Qi, Y.~Zhou, Y.-F. Liu, L.~Liu, and Z.~Pan, ``Traffic-aware task offloading based on convergence of communication and sensing in vehicular edge computing,'' \emph{IEEE Internet Things J.}, vol.~8, no.~24, pp. 17\,762--17\,777, Dec. 2021.

\bibitem{9964436}
Y.~Peng, X.~Tang, Y.~Zhou, J.~Li, Y.~Qi, L.~Liu, and H.~Lin, ``Computing and communication cost-aware service migration enabled by transfer reinforcement learning for dynamic vehicular edge computing networks,'' \emph{IEEE Trans. Mob. Comput.}, vol.~23, no.~1, pp. 257--269, Jan. 2024.

\bibitem{9380744}
X.~Hu, C.~Masouros, and K.-K. Wong, ``Reconfigurable intelligent surface aided mobile edge computing: From optimization-based to location-only learning-based solutions,'' \emph{IEEE Trans. Commun.}, vol.~69, no.~6, pp. 3709--3725, Jun. 2021.

\bibitem{8986845}
F.~Zhou and R.~Q. Hu, ``Computation efficiency maximization in wireless-powered mobile edge computing networks,'' \emph{IEEE Trans. Wireless Commun.}, vol.~19, no.~5, pp. 3170--3184, May 2020.

\bibitem{9018180}
H.~Chen, D.~Zhao, Q.~Chen, and R.~Chai, ``Joint computation offloading and radio resource allocations in small-cell wireless cellular networks,'' \emph{IEEE Trans. Green Commun. Netw.}, vol.~4, no.~3, pp. 745--758, Sep. 2020.

\bibitem{9270605}
Z.~Chu, P.~Xiao, M.~Shojafar, D.~Mi, J.~Mao, and W.~Hao, ``Intelligent reflecting surface assisted mobile edge computing for internet of things,'' \emph{IEEE Wireless Commun. Lett.}, vol.~10, no.~3, pp. 619--623, Mar. 2021.

\bibitem{10120724}
H.~Xie, M.~Xia, P.~Wu, S.~Wang, and H.~V. Poor, ``Edge learning for large-scale internet of things with task-oriented efficient communication,'' \emph{IEEE Trans. Wireless Commun.}, pp. 1--1, Early Access 2023.

\bibitem{4675744}
B.~L. Ng, J.~S. Evans, S.~V. Hanly, and D.~Aktas, ``Distributed downlink beamforming with cooperative base stations,'' \emph{IEEE Trans. Inf. Theory}, vol.~54, no.~12, pp. 5491--5499, Dec. 2008.

\bibitem{8575160}
L.~Liu, Y.~Zhou, W.~Zhuang, J.~Yuan, and L.~Tian, ``Tractable coverage analysis for hexagonal macrocell-based heterogeneous udns with adaptive interference-aware comp,'' \emph{IEEE Trans. Wireless Commun.}, vol.~18, no.~1, pp. 503--517, Jan. 2019.

\bibitem{8110665}
L.~Liu, Y.~Zhou, V.~Garcia, L.~Tian, and J.~Shi, ``Load aware joint comp clustering and inter-cell resource scheduling in heterogeneous ultra dense cellular networks,'' \emph{IEEE Trans. Veh. Technol.}, vol.~67, no.~3, pp. 2741--2755, Mar. 2018.

\bibitem{10091816}
Q.~Cai, Y.~Zhou, L.~Liu, Y.~Qi, Z.~Pan, and H.~Zhang, ``Collaboration of heterogeneous edge computing paradigms: How to fill the gap between theory and practice,'' \emph{IEEE Wireless Commun.}, pp. 1--9, Early Access 2023.

\bibitem{8376975}
J.~T. Chapman, J.~Andreoli-Fang, M.~Chauvin, E.~C. Reyes, Z.~Lu, D.~Liu, J.~Padden, and A.~Bernstein, ``Low latency techniques for mobile backhaul over {DOCSIS},'' in \emph{Proc. IEEE Wireless Commun. Netw. Conf. (WCNC)}, Apr. 2018, pp. 1--6.

\bibitem{6612630}
L.~Su, C.~Yang, and S.~Han, ``The value of channel prediction in {CoMP} systems with large backhaul latency,'' \emph{IEEE Trans. Commun.}, vol.~61, no.~11, pp. 4577--4590, Nov. 2013.

\bibitem{raaen2015measuring}
K.~Raaen and I.~Kjellmo, ``Measuring latency in virtual reality systems,'' in \emph{Proc. Int. Conf. Entertainment Comput.}, 2015, pp. 457--462.

\bibitem{7542156}
Y.~Wang, M.~Sheng, X.~Wang, L.~Wang, and J.~Li, ``Mobile-edge computing: Partial computation offloading using dynamic voltage scaling,'' \emph{IEEE Trans. Commun.}, vol.~64, no.~10, pp. 4268--4282, Oct. 2016.

\bibitem{9950554}
Z.~Chu, P.~Xiao, M.~Shojafar, D.~Mi, W.~Hao, J.~Shi, and F.~Zhou, ``Utility maximization for {IRS} assisted wireless powered mobile edge computing and caching {(WP-MECC)} networks,'' \emph{IEEE Trans. Commun.}, vol.~71, no.~1, pp. 457--472, Jan. 2023.

\bibitem{9062599}
K.~Li, M.~Tao, and Z.~Chen, ``Exploiting computation replication for mobile edge computing: A fundamental computation-communication tradeoff study,'' \emph{IEEE Trans. Wireless Commun.}, vol.~19, no.~7, pp. 4563--4578, Jul. 2020.

\bibitem{9087848}
Q.-U.-A. Nadeem, H.~Alwazani, A.~Kammoun, A.~Chaaban, M.~Debbah, and M.-S. Alouini, ``Intelligent reflecting surface-assisted multi-user {MISO} communication: Channel estimation and beamforming design,'' \emph{IEEE Open J. Commun. Soc.}, vol.~1, pp. 661--680, May 2020.

\bibitem{9115725}
S.~Abeywickrama, R.~Zhang, Q.~Wu, and C.~Yuen, ``Intelligent reflecting surface: Practical phase shift model and beamforming optimization,'' \emph{IEEE Trans. Commun.}, vol.~68, no.~9, pp. 5849--5863, Sep. 2020.

\bibitem{9551980}
X.~Pei, H.~Yin, L.~Tan, L.~Cao, Z.~Li, K.~Wang, K.~Zhang, and E.~Björnson, ``{RIS-Aided} wireless communications: Prototyping, adaptive beamforming, and indoor/outdoor field trials,'' \emph{IEEE Trans. Commun.}, vol.~69, no.~12, pp. 8627--8640, Dec. 2021.

\bibitem{ericsson2023}
``5{G} {NR} testbed 3.5 {GH}z coverage results,'' \url{https://www.ericsson.com/en/reports-and-papers/research-papers/5g-nr-testbed-3.5-ghz-coverage-results}, 2018, accessed: August 2, 2023.

\bibitem{qualcomm2023}
``White paper: 5{G} {NR} millimeter wave network coverage simulation,'' \url{https://www.qualcomm.com/content/dam/qcomm-martech/dm-assets/documents/5g_nr_millimeter_wave_network_coverage_simulation _studies_for_global_cities.pdf}, 2017, accessed: August 2, 2023.

\bibitem{9352968}
S.~Hua, Y.~Zhou, K.~Yang, Y.~Shi, and K.~Wang, ``Reconfigurable intelligent surface for green edge inference,'' \emph{IEEE Trans. Green Commun. Netw.}, vol.~5, no.~2, pp. 964--979, 2021.

\bibitem{10221786}
W.~He, D.~He, X.~Ma, X.~Chen, Y.~Fang, and W.~Zhang, ``Joint user association, resource allocation, and beamforming in ris-assisted multi-server mec systems,'' \emph{IEEE Trans. Wireless Commun.}, pp. 1--1, Early Access 2023.

\bibitem{9570143}
X.~Mu, Y.~Liu, L.~Guo, J.~Lin, and R.~Schober, ``Simultaneously transmitting and reflecting ({STAR}) {RIS} aided wireless communications,'' \emph{IEEE Trans. Wireless Commun.}, vol.~21, no.~5, pp. 3083--3098, May 2022.

\bibitem{6574874}
W.~Zhang, Y.~Wen, K.~Guan, D.~Kilper, H.~Luo, and D.~O. Wu, ``Energy-optimal mobile cloud computing under stochastic wireless channel,'' \emph{IEEE Trans. Wireless Commun.}, vol.~12, no.~9, pp. 4569--4581, Sep. 2013.

\bibitem{8334188}
S.~Bi and Y.~J. Zhang, ``Computation rate maximization for wireless powered mobile-edge computing with binary computation offloading,'' \emph{IEEE Trans. Wireless Commun.}, vol.~17, no.~6, pp. 4177--4190, Jun. 2018.

\bibitem{9812481}
Y.~Ye, L.~Shi, X.~Chu, R.~Q. Hu, and G.~Lu, ``Resource allocation in backscatter-assisted wireless powered {MEC} networks with limited {MEC} computation capacity,'' \emph{IEEE Trans. Wireless Commun.}, vol.~21, no.~12, pp. 10\,678--10\,694, Dec. 2022.

\bibitem{yang2020sparse}
X.~Yang, S.~Hua, Y.~Shi, H.~Wang, J.~Zhang, and K.~B. Letaief, ``Sparse optimization for green edge {AI} inference,'' \emph{J. Commun. Inf. Netw.}, vol.~5, no.~1, pp. 1--15, Mar. 2020.

\bibitem{lovasz1995randomized}
L.~Lovasz, ``Randomized algorithms in combinatorial optimization,'' \emph{Combinatorial Optimization, volume 20 of DIMACS Series in Discrete Mathematics and Theoretical Computer Science}, vol.~20, pp. 153--179, 1995.

\bibitem{neumann2006combinatorial}
F.~Neumann, ``Combinatorial optimization and the analysis of randomized search heuristics,'' Ph.D. dissertation, Christian-Albrechts Universit{\"a}t Kiel, 2006.

\bibitem{9234651}
C.~Park and J.~Lee, ``Mobile edge computing-enabled heterogeneous networks,'' \emph{IEEE Trans. Wireless Commun.}, vol.~20, no.~2, pp. 1038--1051, Feb. 2021.

\bibitem{9286485}
Y.~Qu, H.~Dai, F.~Wu, D.~Lu, C.~Dong, S.~Tang, and G.~Chen, ``Robust offloading scheduling for mobile edge computing,'' \emph{IEEE Trans. Mob. Comput.}, vol.~21, no.~7, pp. 2581--2595, Jul. 2022.

\bibitem{9031741}
K.~Yang, Y.~Shi, W.~Yu, and Z.~Ding, ``Energy-efficient processing and robust wireless cooperative transmission for edge inference,'' \emph{IEEE Internet Things J.}, vol.~7, no.~10, pp. 9456--9470, Oct. 2020.

\bibitem{visotsky1999optimum}
E.~Visotsky and U.~Madhow, ``Optimum beamforming using transmit antenna arrays,'' in \emph{Proc. IEEE Veh. Technol. Conf}, vol.~1, May 1999, pp. 851--856.

\bibitem{9435051}
L.~Du, S.~Shao, G.~Yang, J.~Ma, Q.~Liang, and Y.~Tang, ``Capacity characterization for reconfigurable intelligent surfaces assisted multiple-antenna multicast,'' \emph{IEEE Trans. Wireless Commun.}, vol.~20, no.~10, pp. 6940--6953, Oct. 2021.

\bibitem{10121446}
Z.~Liu, Z.~Li, M.~Wen, Y.~Gong, and Y.-C. Wu, ``{STAR}-{RIS}-aided mobile edge computing: Computation rate maximization with binary amplitude coefficients,'' \emph{IEEE Trans. Commun.}, vol.~71, no.~7, pp. 4313--4327, Jul. 2023.

\bibitem{andresen2016convergence}
A.~Andresen and V.~Spokoiny, ``Convergence of an alternating maximization procedure,'' \emph{The J. Mach. Learn. Res.}, vol.~17, no.~1, pp. 2229--2281, Apr. 2016.

\bibitem{6832894}
E.~Björnson, M.~Bengtsson, and B.~Ottersten, ``Optimal multiuser transmit beamforming: A difficult problem with a simple solution structure [lecture notes],'' \emph{IEEE Signal Process. Mag.}, vol.~31, no.~4, pp. 142--148, Jul. 2014.

\bibitem{seneviratne2012l0}
S.~Akila, ``$l_0$ sparse signal processing and model selection with applications,'' Ph.D. dissertation, UNSW Sydney, 2012.

\bibitem{polik2010interior}
I.~P{\'o}lik and T.~Terlaky, ``Interior point methods for nonlinear optimization,'' in \emph{Nonlinear optimization}.\hskip 1em plus 0.5em minus 0.4em\relax Springer, Berlin, Heidelberg, 2010, pp. 215--276.

\bibitem{1255564}
R.~Gribonval and M.~Nielsen, ``Sparse representations in unions of bases,'' \emph{IEEE Trans. Inf. Theory}, vol.~49, no.~12, pp. 3320--3325, Dec. 2003.

\bibitem{li2022phase}
Z.~Li, S.~Wang, Q.~Lin, Y.~Li, M.~Wen, Y.-C. Wu, and H.~V. Poor, ``Phase shift design in {RIS} empowered wireless networks: from optimization to {AI}-based methods,'' \emph{Network}, vol.~2, no.~3, pp. 398--418, 2022.

\bibitem{8811733}
Q.~Wu and R.~Zhang, ``Intelligent reflecting surface enhanced wireless network via joint active and passive beamforming,'' \emph{IEEE Trans. Wireless Commun.}, vol.~18, no.~11, pp. 5394--5409, 2019.

\bibitem{8741198}
C.~Huang, A.~Zappone, G.~C. Alexandropoulos, M.~Debbah, and C.~Yuen, ``Reconfigurable intelligent surfaces for energy efficiency in wireless communication,'' \emph{IEEE Trans. Wireless Commun.}, vol.~18, no.~8, pp. 4157--4170, Nov. 2019.

\bibitem{9457078}
Y.~Chen, M.~Wen, E.~Basar, Y.-C. Wu, L.~Wang, and W.~Liu, ``Exploiting reconfigurable intelligent surfaces in edge caching: Joint hybrid beamforming and content placement optimization,'' \emph{IEEE Trans. Wireless Commun.}, vol.~20, no.~12, pp. 7799--7812, Dec. 2021.

\bibitem{heath2018scientific}
M.~T. Heath, \emph{Scientific computing: an introductory survey, revised second edition}.\hskip 1em plus 0.5em minus 0.4em\relax Philadelphia, PA, USA: SIAM, 2018.

\bibitem{boyd2004convex}
S.~Boyd and L.~Vandenberghe, \emph{Convex optimization}.\hskip 1em plus 0.5em minus 0.4em\relax Cambridge, U.K.: Cambridge Univ. Press, 2004.

\bibitem{8410057}
Y.~Li, M.~Xia, and Y.-C. Wu, ``First-order algorithm for content-centric sparse multicast beamforming in large-scale {C-RAN},'' \emph{IEEE Trans. Wireless Commun.}, vol.~17, no.~9, pp. 5959--5974, Sep. 2018.

\bibitem{lipp2016variations}
T.~Lipp and S.~Boyd, ``Variations and extension of the convex-concave procedure,'' \emph{Optim. Eng.}, vol.~17, pp. 263--287, 2016.

\bibitem{8474356}
K.-G. Nguyen, Q.-D. Vu, L.-N. Tran, and M.~Juntti, ``Energy efficiency fairness for multi-pair wireless-powered relaying systems,'' \emph{IEEE J. Sel. Areas Commun.}, vol.~37, no.~2, pp. 357--373, Feb. 2019.

\bibitem{5447068}
Z.-q. Luo, W.-k. Ma, A.~M.-c. So, Y.~Ye, and S.~Zhang, ``Semidefinite relaxation of quadratic optimization problems,'' \emph{IEEE Signal Process. Mag.}, vol.~27, no.~3, pp. 20--34, May 2010.

\bibitem{tse2005fundamentals}
D.~Tse and P.~Viswanath, \emph{Fundamentals of wireless communication}.\hskip 1em plus 0.5em minus 0.4em\relax Cambridge, U.K.: Cambridge Univ. Press, 2005.

\bibitem{8941080}
Q.~Wu and R.~Zhang, ``Weighted sum power maximization for intelligent reflecting surface aided {SWIPT},'' \emph{IEEE Wireless Commun. Lett.}, vol.~9, no.~5, pp. 586--590, May 2020.

\bibitem{9133107}
T.~Bai, C.~Pan, Y.~Deng, M.~Elkashlan, A.~Nallanathan, and L.~Hanzo, ``Latency minimization for intelligent reflecting surface aided mobile edge computing,'' \emph{IEEE J. Sel. Areas Commun.}, vol.~38, no.~11, pp. 2666--2682, Nov. 2020.

\bibitem{9764640}
R.~Pinto~Antonioli, I.~M. Braga, G.~Fodor, Y.~C.~B. Silva, A.~L.~F. de~Almeida, and W.~C. Freitas, ``On the energy efficiency of {Cell-Free} systems with limited fronthauls: Is coherent transmission always the best alternative?'' \emph{IEEE Trans. Wireless Commun.}, vol.~21, no.~10, pp. 8729--8743, Oct. 2022.

\bibitem{8097026}
H.~Q. Ngo, L.-N. Tran, T.~Q. Duong, M.~Matthaiou, and E.~G. Larsson, ``On the total energy efficiency of {Cell-Free} massive {MIMO},'' \emph{IEEE Trans. Green Commun. Netw.}, vol.~2, no.~1, pp. 25--39, Mar. 2018.

\end{thebibliography}

\vspace{11pt}

\vfill
\end{document}